\documentclass[12pt]{article}
\usepackage{amsmath,amsfonts,amssymb}
\usepackage{hyperref}
\usepackage{graphicx}
\usepackage[dvips]{color}
\usepackage{epsfig}
\usepackage{psfrag}
\allowdisplaybreaks

\usepackage{color}

\newcommand \be{\begin{eqnarray}}
\newcommand \ee{\end{eqnarray}}

\makeatletter\@addtoreset{equation}{section}\makeatother

\DeclareMathOperator{\Tr}{Tr}

\DeclareMathOperator{\Li}{Li}

\DeclareMathOperator{\diag}{diag}
\DeclareMathOperator{\sign}{sign}

\DeclareMathOperator{\sn}{sn}
\DeclareMathOperator{\cn}{cn}
\DeclareMathOperator{\dn}{dn}
\DeclareMathOperator{\am}{am}
\DeclareMathOperator{\ns}{ns}

\DeclareMathOperator{\nd}{nd}
\DeclareMathOperator{\sd}{sd}
\DeclareMathOperator{\cd}{cd}

\def\bR {\mathbb{R}}
\def\bS {\mathbb{S}}

\def \EE {{\mathbb{E}}}
\def \KK {{\mathbb{K}}}

\newcommand{\beq}{\begin{equation}}
\newcommand{\eeq}{\end{equation}}
\newcommand{\bal}{\begin{equation}\begin{aligned}}
\newcommand{\eal}{\end{aligned} \end{equation}}
\newcommand{\bea}{\begin{eqnarray}}
\newcommand{\eea}{\end{eqnarray}}

\newcommand{\vev}[1]{{\left< {#1} \right>}}

\newcommand{\eqn}[1]{(\ref{#1})}

\newcommand{\address}[1]{\vbox{\center\em#1}}
\renewcommand{\title}[1]{\vbox{\center\huge{#1}}\vspace{5mm}}

\newcommand{\cL}{{\mathcal L}}

\newcommand{\cN}{{\mathcal N}}
\newcommand{\cP}{{\mathcal P}}

\newcommand{\cO}{{\mathcal O}}
\newcommand{\cS}{{\mathcal S}}

\begin{document}
\bibliographystyle{utphys}

\begin{titlepage}
\begin{center}
\vskip5mm

\hfill {\tt Imperial-TP-2011-ND-02}\\
\hfill {\tt NSF-KITP-11-073}\\
\hfill {\tt AEI-2011-027}

\includegraphics[width=20mm]{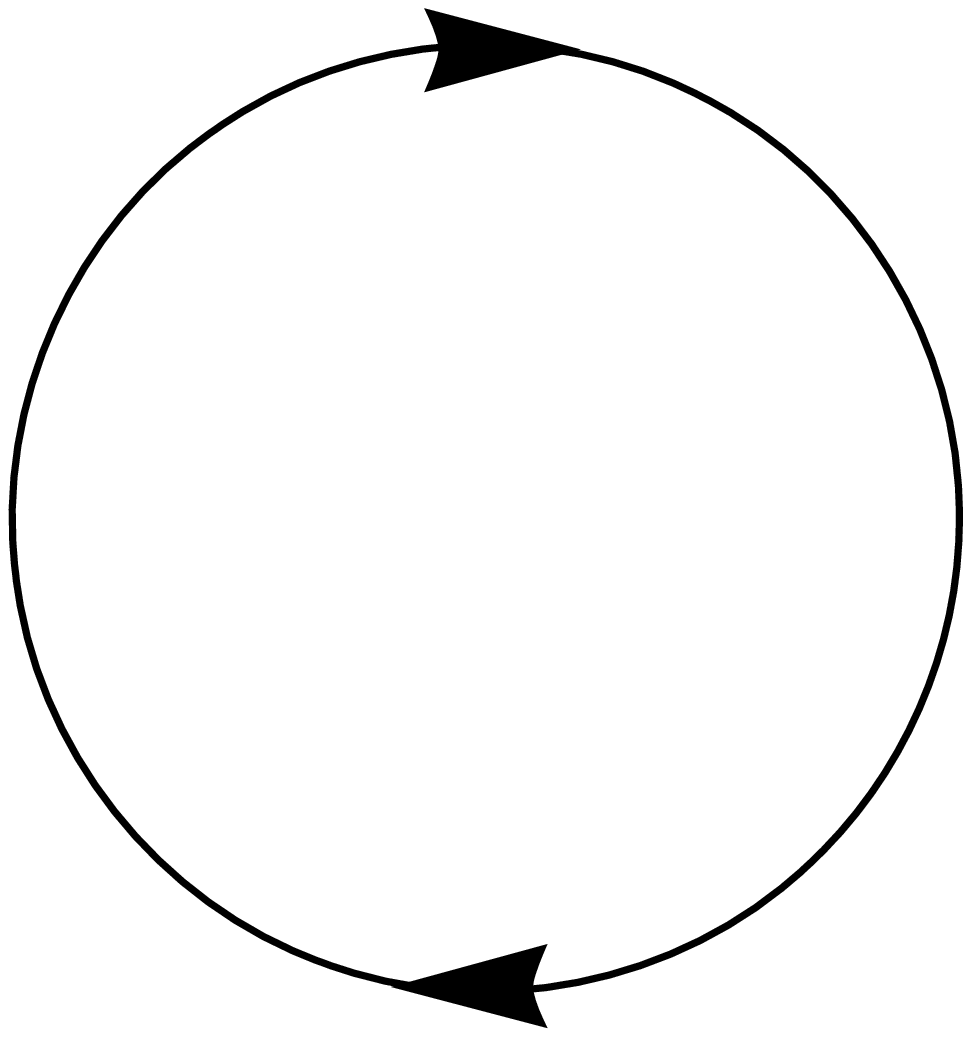}
\qquad
\raisebox{8mm}{\text{\huge{$\to$}}}
\qquad
\includegraphics[width=30mm]{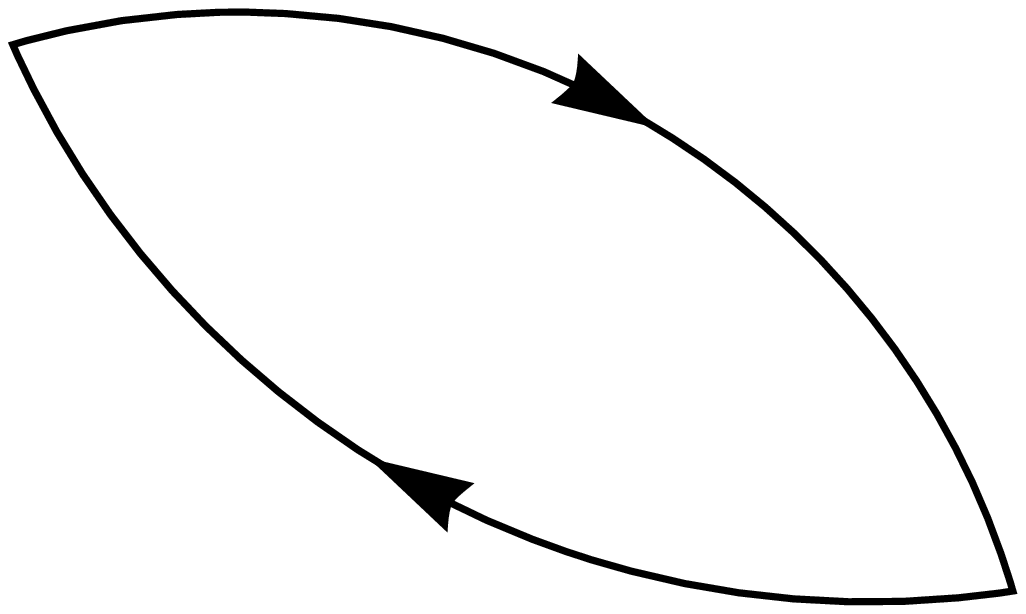}
\qquad
\raisebox{8mm}{\text{\huge{$\to$}}}
\qquad
\raisebox{2mm}{\includegraphics[width=25mm]{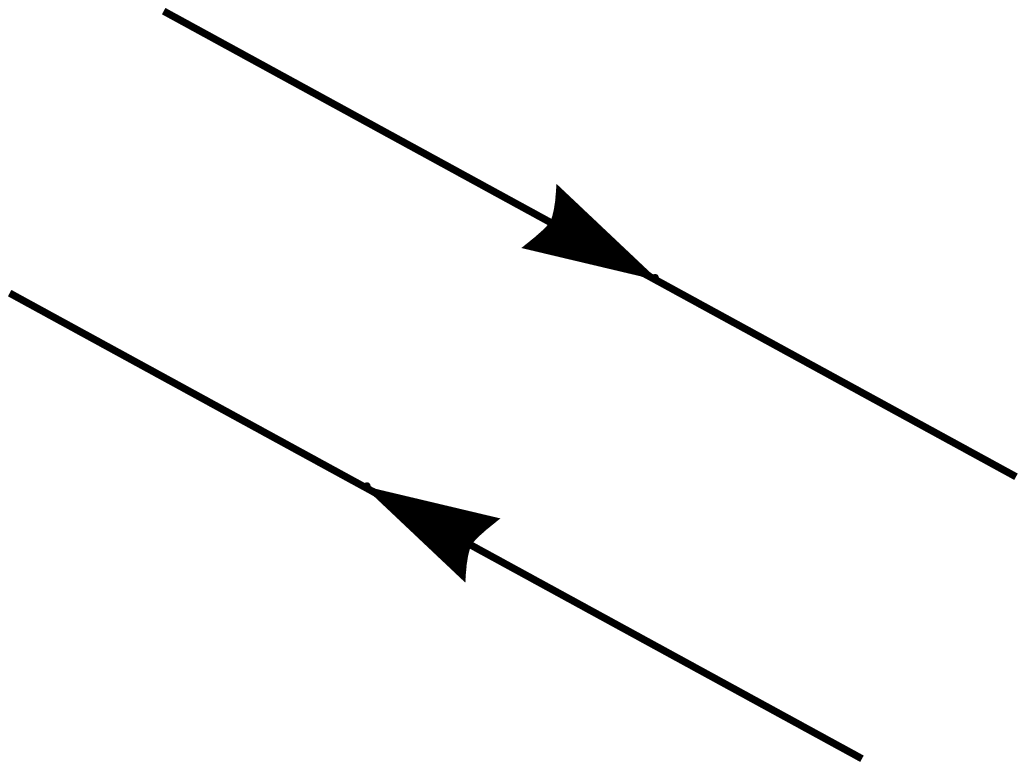}}

\vspace{2mm}

\title{Generalized quark-antiquark potential\\
at weak and strong coupling}
\vspace{5mm}

\renewcommand{\thefootnote}{$\alph{footnote}$}

Nadav Drukker\footnote{\href{mailto:ndrukker@imperial.ac.uk}
{\tt ndrukker@imperial.ac.uk}} and
Valentina Forini\footnote{\href{mailto:forini@aei.mpg.de}
{\tt forini@aei.mpg.de}}
\vskip 5mm
\address{
${}^{a}$The Blackett Laboratory, Imperial College London,\\
Prince Consort Road, London SW7 2AZ, U.K.\\
}
\address{
${}^{b}$Max-Planck-Institut f\"ur Gravitationsphysik, Albert-Einstein-Institut, \\
Am M\"uhlenberg 1, D-14476 Potsdam, Germany 
}

\renewcommand{\thefootnote}{\arabic{footnote}}
\setcounter{footnote}{0}

\end{center}

\abstract{
\noindent
We study a two--parameter family of Wilson loop operators in $\cN=4$ supersymmetric 
Yang-Mills theory which interpolates smoothly between the $1/2$ BPS line or circle 
and a pair of antiparallel lines. These observables capture a natural generalization of the 
quark-antiquark potential. We calculate these loops on the gauge theory side to 
second order in perturbation theory and in a semiclassical expansion in string theory 
to one--loop order. The resulting determinants are given in integral form and can 
be evaluated numerically for general values of the parameters or analytically 
in a systematic expansion around the $1/2$ BPS configuration. 
We comment about the feasibility of deriving all--loop results for these Wilson loops.
}

\begin{center}
\includegraphics[width=33mm]{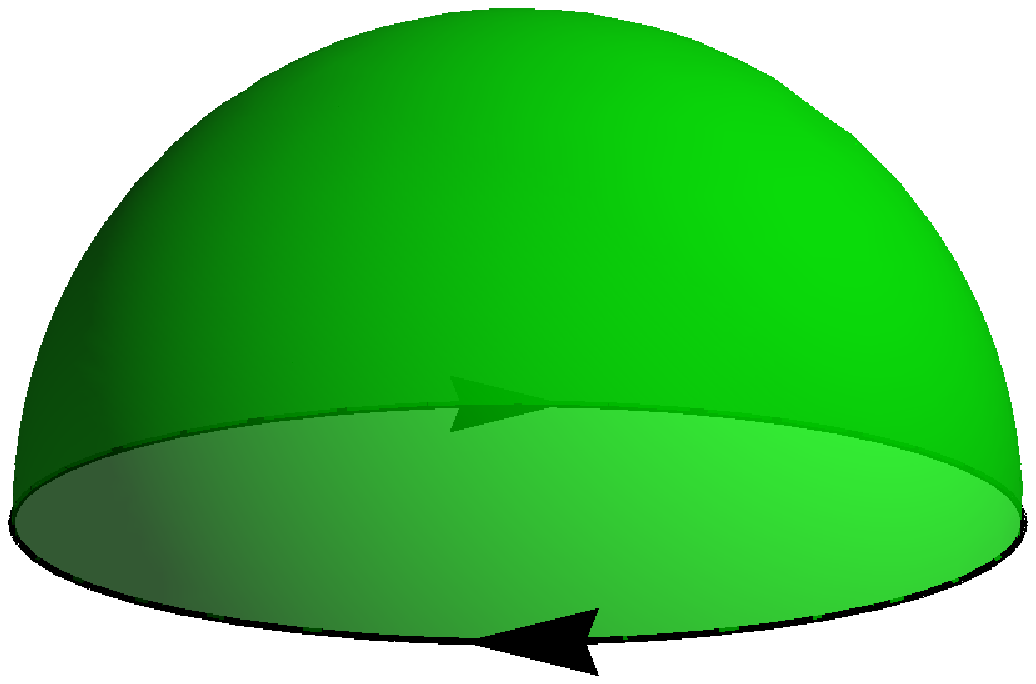}
\quad
\raisebox{8mm}{\text{\huge{$\to$}}}
\raisebox{-2mm}{\includegraphics[width=45mm]{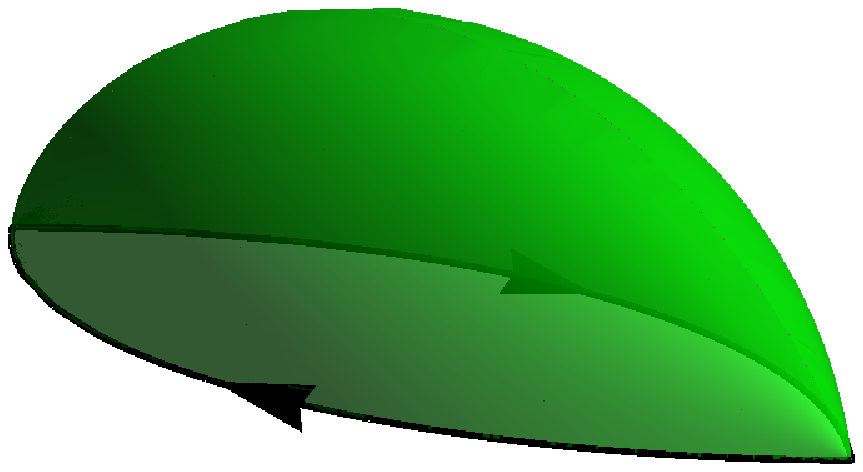}}
\raisebox{8mm}{\text{\huge{$\to$}}}
\quad
\includegraphics[width=35mm]{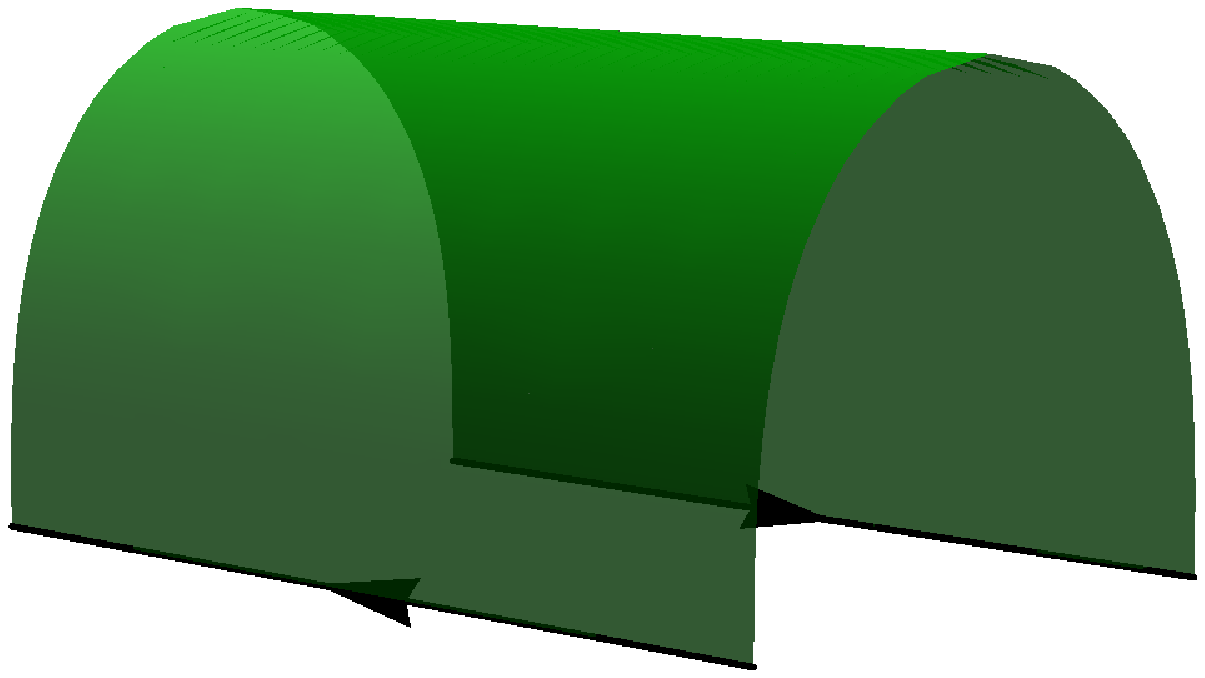}

\end{center}

\end{titlepage}

\setcounter{tocdepth}{1}
{\addtolength{\parskip}{-1.7ex}
\tableofcontents
}

\section{Introduction}
\label{sec:intro}

The duality between $\cN=4$ SYM in four dimensions and type IIB string theory on 
$AdS_5\times\bS^5$ has gone through a revolution in the past few years. From 
the early study of general features, protected quantities and BPS observables, 
it evolved into the precision study of the spectrum of local operators and of scattering 
amplitudes. In both these examples the (conjectured) integrability of the planar 
theory enabled great calculational leaps allowing in certain cases to find 
non-trivial interpolating functions matching all known explicit weak and 
strong coupling tests as well as satisfying very restrictive consistency conditions.

A natural family of observables in the gauge theory are Wilson loops, realized 
in the dual theory as infinite open strings. When the asymptotic boundary is 
along light--like segments, they are related by T-duality to scattering amplitudes 
\cite{am1}. The most natural Wilson loop, though, is a pair of antiparallel lines 
in flat space, which calculates the effective potential between a pair of 
infinitely heavy W-boson probes.

The expectation value of this observable was calculated very early after the 
introduction of the $AdS$/CFT duality by a classical string configuration 
\cite{maldacena-wl, rey-yee}. Since then 
very little progress has been made in understanding this quantity. It has been 
calculated at weak coupling to second order in perturbation theory \cite{essz,esz,Pineda} 
and on the string theory side the problem of calculating the first correction 
(of relative order $\lambda^{-1/2}$) was formulated in \cite{fgt,dgt}, 
evaluated numerically in \cite{chr} and simplified further in~\cite{vali-lines} to 
an analytic one--dimensional integral.

As it turned out, the circular Wilson loop is a much simpler observable, which 
in the Feynman gauge receives contributions only from ladder graphs 
\cite{esz, dg-mm,pestun}. For the 
antiparallel lines the ladder graphs give an answer of the same order as that 
calculated by string theory, but not identical \cite{essz,esz}.

We examine here this problem again and propose a program which may allow for 
an exact calculation of the expectation value of such unprotected Wilson loops at 
all values of the coupling.

Since $\cN=4$ SYM is conformal, the potential calculated by the Wilson loop 
has the Coulomb form with a coefficient which is coupling dependent. It would 
be very useful to introduce extra parameters to the problem which one could 
vary to get a handle on the calculation. In fact there are two simple 
deformations of the problem which do not make the perturbative or supergravity 
calculation any harder and allow to interpolate between protected operators and 
the desired observable.

The first deformation parameter was introduced already in \cite{maldacena-wl}, 
and allows for the two lines to couple to two different scalar fields. We label this 
parameter $\theta$. For $\theta=0$ the two lines couple to the same scalar 
field, say $\Phi_1$. When $\theta=\pi/2$ the two lines couple to 
$\Phi_1\pm\Phi_2$, which are orthogonal to each-other. 
Then for $\theta=\pi$ they couple to the field $\Phi_2$, but with opposite 
signs, which means that the lines are effectively parallel, rather than antiparallel. 
In that case the two lines share eight supercharges and the correlator is trivial.

The other deformation parameter is geometric. One way to illustrate it is to replace 
the theory on $\bR^4$ with the theory on $\bS^3\times\bR$ (related by the exponential 
map). Now we consider 
a pair of antiparallel lines separated by an angle $\pi-\phi$ on $\bS^3$. For 
$\phi=0$ the two lines are antipodal and mutually BPS, while for $\phi\to\pi$ the 
lines get very close together. If we ``zoom in'' to the vicinity of the lines by a 
conformal transformation we get a situation very similar to the original antiparallel 
lines in flat space.

Different points of view on this deformation are presented in the following section, 
and the $\phi\to\pi$ limit is explored in more detail in Section~\ref{sec:anti}. 
Let us note here 
only that an equivalent picture is that of a cusp in the plane in $\bR^4$. For $\phi=0$ 
the cusp disappears and the system is that of a single infinite straight line.

Going back to the $\bS^3\times\bR$ picture, the expectation value of the Wilson 
loop calculates the effective potential $V(\phi,\theta,\lambda)$ between a generalized 
quark antiquark pair, in exactly the same way as originally proposed by Wilson 
\cite{wilson}. The operator is made of a pair of lines extending over a large time 
$T$ and can be written as
\beq
W=\frac{1}{N}\Tr\cP\exp\left[\oint (iA_\mu\dot x^\mu+\Phi_I\Theta^I|\dot x|)ds\right].
\label{wl}
\eeq
The expectation value of the loop operator has the behavior
\beq
\vev{W}\approx \exp\Big[{-}T\,V(\phi,\theta,\lambda)\Big]\,.
\label{2-lines-vev}
\eeq

The effective potential $V(\phi,\theta,\lambda)$ depends on the 't~Hooft coupling 
$\lambda=g^2N$ (we do not consider non-planar corrections) and it can be 
expanded at weak coupling in a perturbative series
\beq
V(\phi,\theta,\lambda)=\sum_{n=1}^\infty \left(\frac{\lambda}{16\pi^2}\right)^nV^{(n)}(\phi,\theta)\,.
\eeq
In section~\ref{sec:pert} we present the exact form of the first two terms in this expansion, 
based on \cite{mos}.

The strong coupling behavior of $V(\phi,\theta,\lambda)$ can be calculated in a systematic 
expansion around a classical string solution. The function is expected to have an asymptotic 
expansion of the form
\beq
V(\phi,\theta,\lambda)=\frac{\sqrt\lambda}{4\pi}
\sum_{l=0}^\infty \left(\frac{4\pi}{\sqrt\lambda}\right)^lV_{AdS}^{(l)}(\phi,\theta)\,.
\label{ads-series}
\eeq
In Section~\ref{sec:ads} we study this expansion. $V_{AdS}^{(0)}$ is proportional to the 
classical action of the string solution, calculated originally in \cite{dgo,dgrt-big}. The next 
term, $V_{AdS}^{(1)}$, requires tracing over all fluctuation modes and we derive an 
integral expression which can be evaluated numerically to high precision 
for fixed $\phi$ and $\theta$.

The coefficients in the perturbative expansions are complicated functions of the 
angles $\phi$ and $\theta$ and at strong coupling they are given only implicitly 
(at the classical level) or in integral form (one--loop). We consider therefore 
the expansion of these functions around $\phi=\theta=0$. This is an expansion 
around the $1/2$ BPS line (or circle), one of the most simple observables in 
the theory. We view the general problem as a deformation of this $1/2$ BPS 
configuration and use the fact that a deformation of a Wilson loop can be written 
in terms of insertions of local operators into the loop.

Changing $\phi$, which modifies the path of the loop is captured by insertions 
of the field strength $F_{\mu\nu}$, as well as its derivatives, into the Wilson loop. 
Somewhat simpler is to change $\theta$ which introduces local scalar field 
insertions into the loop. Both these quantities can be calculated perturbatively, 
the relevant graphs come from limits of the graphs calculating the effective potential 
at finite values of $\phi$ and $\theta$. On the string side we are also able to 
get exact analytical results for the expansion coefficients, by inverting certain 
transcendental functions perturbatively and from the expansion of elliptic functions 
at small modulus in terms of trigonometric functions, which simplifies the integrals 
arising at one--loop.

In Section~\ref{sec:expand} we present these expansions and explore some 
of their properties. In particular, we can identify specific diagrams in perturbation 
theory contributing to the different terms in the expansion. 

We end with a discussion of our results.

For the benefit of the casual reader, 
we have tried to keep the body of the paper focused on presenting and analyzing our 
results. The derivation of these results is presented in many appendices, dedicated to 
perturbation theory, classical string calculations, the one--loop determinants and their 
various limits.

\section{Setup}
\label{sec:setup}

We would like to present more details about the different ways we can view 
the Wilson loop observables we will study in the rest of the paper.

\begin{figure}[t]
\begin{center}
\includegraphics[height=80mm]{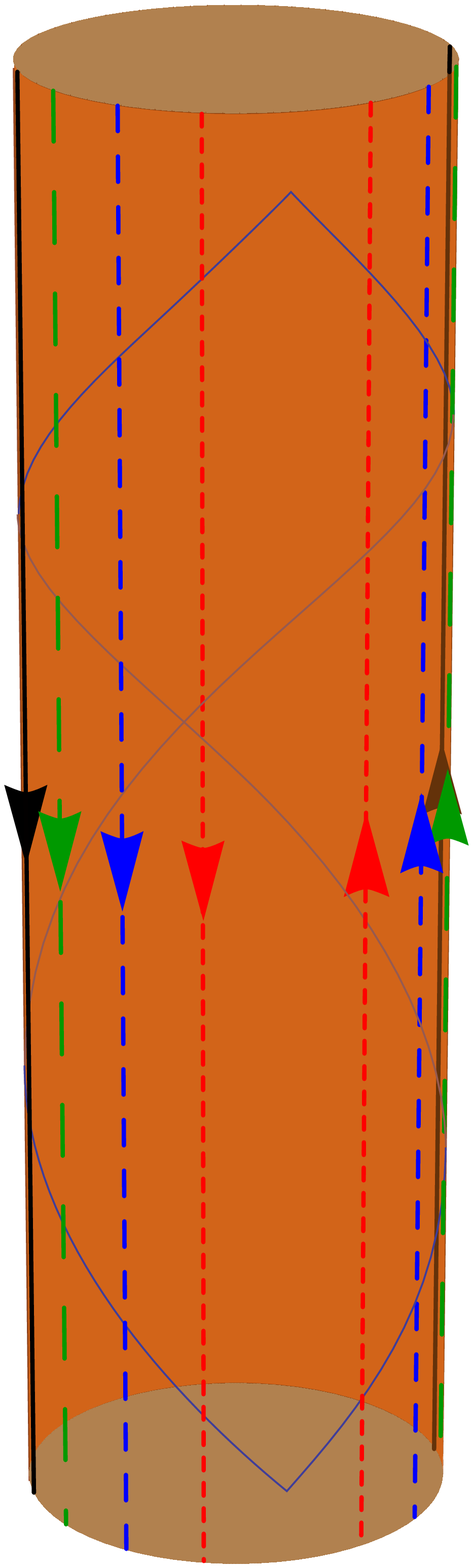}
\qquad
\raisebox{15mm}{
\epsfig{file= 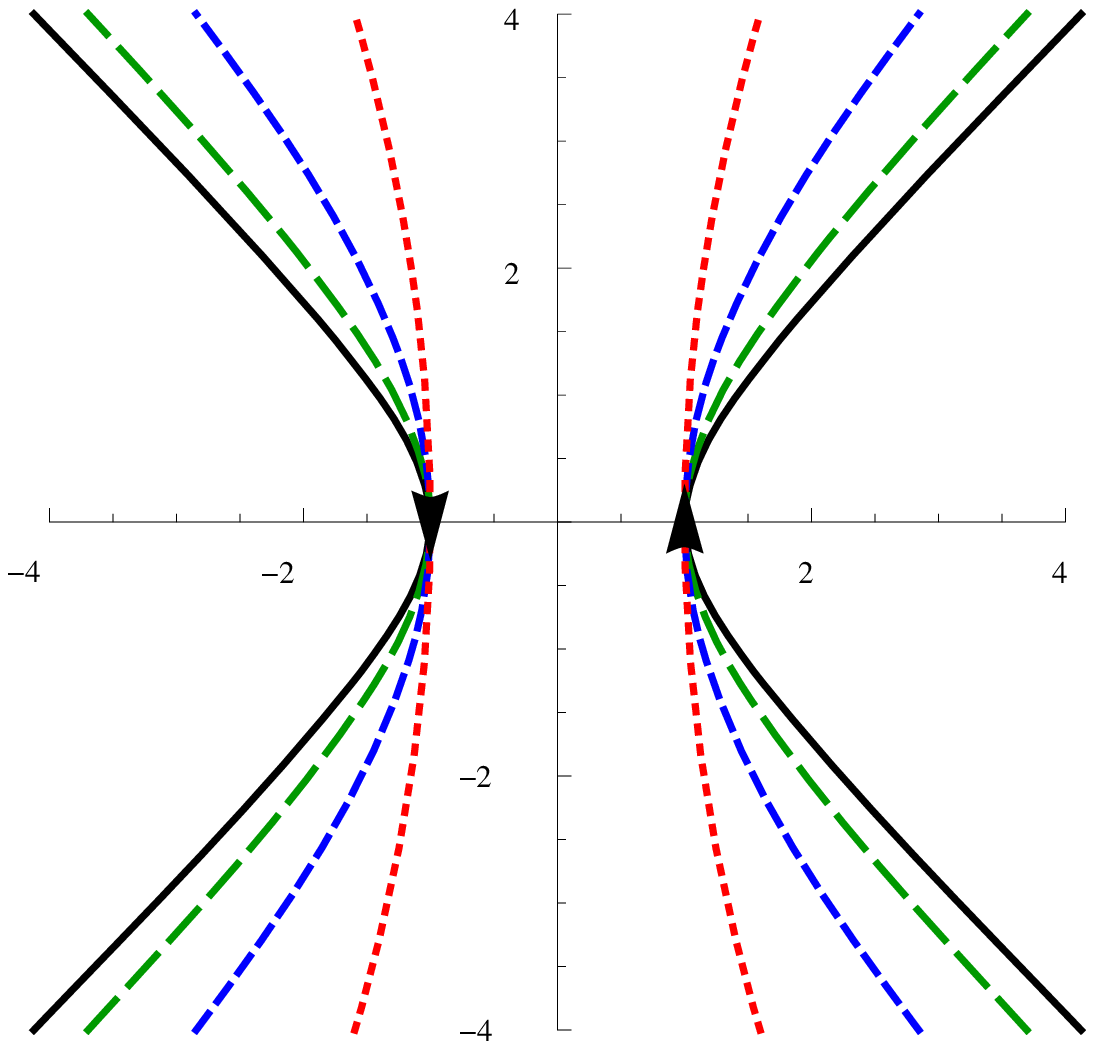,height=60mm
\psfrag{-2}{\footnotesize $-2$}
\psfrag{-4}{\footnotesize $-4$}
\psfrag{2}{\footnotesize $2$}
\psfrag{4}{\footnotesize 4}
}} 
\parbox{13cm}{
\caption{Antiparallel lines on $\bS^3\times\bR$ with Lorentzian signature can 
be mapped by different conformal transformations to hyperbolas in Minkowski space, 
arranged so that they all pass through the points $\pm1$ on the horisontal axis. 
The thin lines on the cylinder map to the boundary of Minkowski space.
\label{fig:hyperbolas}}}
\end{center}
\end{figure}

We start by considering the gauge theory on $\bS^3\times\bR$. The loop is 
made of two lines separated by an angle $\pi-\phi$ along a big circle on $\bS^3$. 
Parameterizing the angle along this circle by $\varphi$ and the time direction 
by $t$ we have a pairs of lines, one going in the future direction and one to 
the past. The parameters appearing in the Wilson loop \eqn{wl} are therefore
\bal
t&=s\,,
&\qquad
\varphi&=\frac{\phi}{2}\,,
&\qquad
\Theta^1&=\cos\frac{\theta}{2}\,,
&\qquad
\Theta^2&=\sin\frac{\theta}{2}\,,\\
t&=-s'\,,
&\qquad
\varphi&=\pi-\frac{\phi}{2}\,,
&\qquad
\Theta^1&=\cos\frac{\theta}{2}\,,
&\qquad
\Theta^2&=-\sin\frac{\theta}{2}\,.
\eal

It is natural (in particular in the context of $AdS$) to use Lorentzian signature on 
this space. A conformal transformation maps a region of $\bS^3\times\bR$ 
to the entire 4d Minkowski space. A straight time--like line gets mapped 
under this transformation to a hyperbola, in the same way that in Euclidean space 
lines get mapped to circles. A Wilson line along such a curve is $1/2$ BPS 
\cite{volker}. The same is true for a pair of hyperbolas sharing the same focal 
point, as they are the image of two antipodal 
lines on the $\bS^3\times\bR$. In all other cases the conformal transformation to Minkowski 
space will give a pair of hyperbola which do not satisfy this property and are not 
mutually BPS. In the limit of $\phi\to\pi$, where the separation of the two lines on 
the cylinder is $\pi-\phi$, the two hyperbola look at the vicinity of the origin like 
two antiparallel lines. See Figure~\ref{fig:hyperbolas}.

If we Wick-rotate $\bS^3\times\bR$ to Euclidean signature, then we can use the 
exponential map to get flat Euclidean space. The pairs of lines running along the time 
direction get mapped to rays intersecting at the origin. The angle between the rays is 
$\pi-\phi$, such that for $\phi=0$ they form a continuous straight line. Otherwise there 
is a singular point.%
\footnote{We propagate the misnomer referring to these Wilson loops as 
having a cusp, even though the singularity has a finite angle.}

\begin{figure}[t]
\begin{center}
\epsfig{file=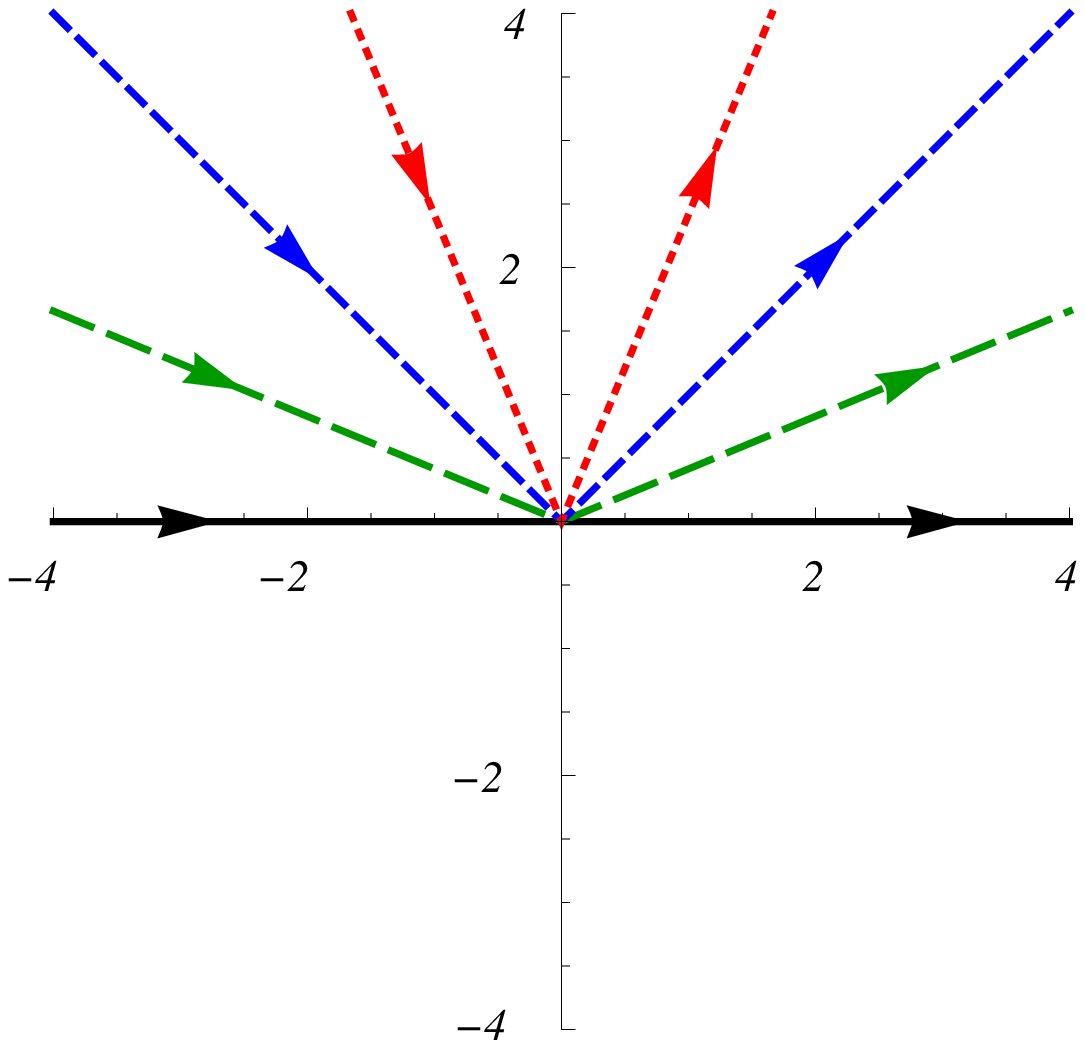,height=60mm
\psfrag{-2}{\footnotesize $-2$}
\psfrag{-4}{\footnotesize $-4$}
\psfrag{2}{\footnotesize $2$}
\psfrag{4}{\footnotesize 4}
} 
\quad
\epsfig{file=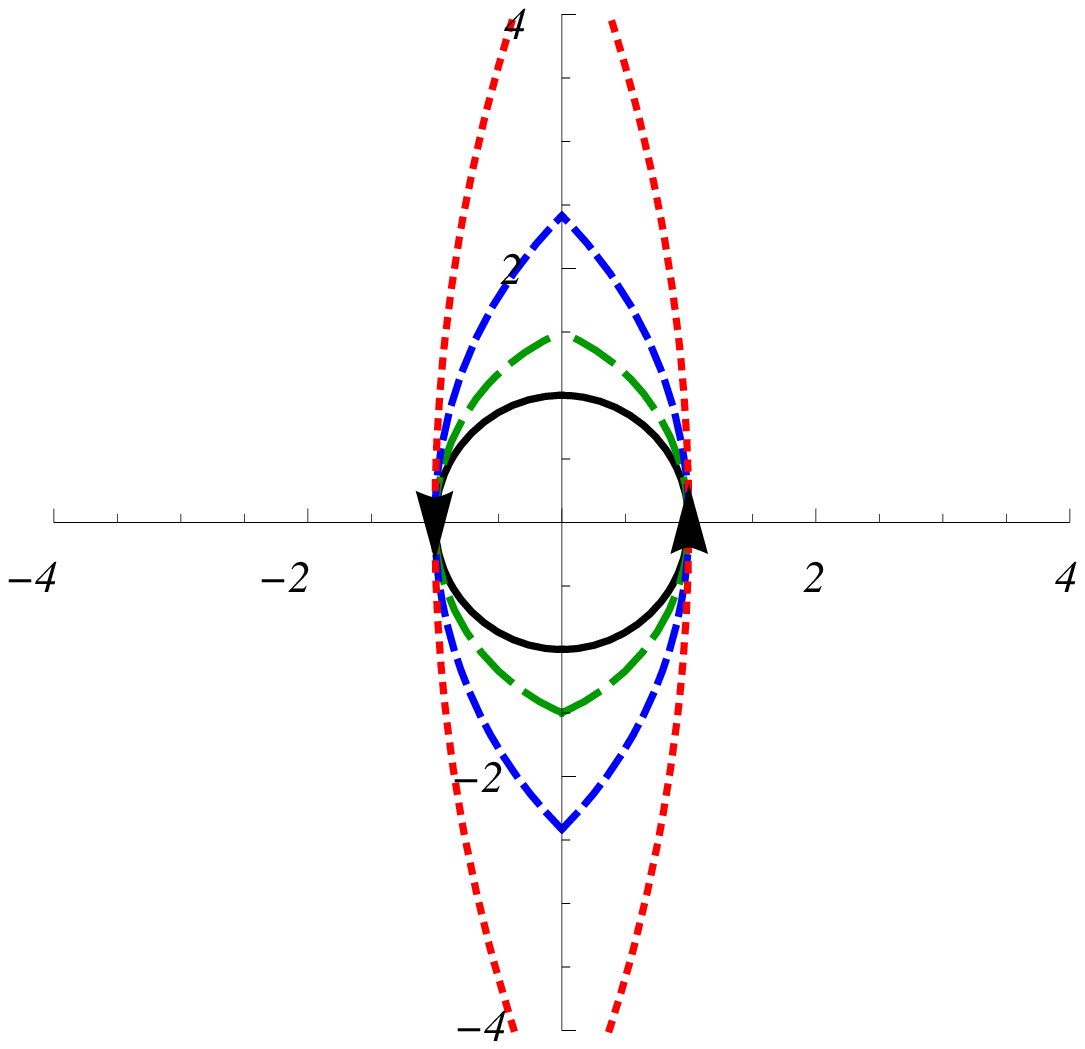,height=60mm
\psfrag{-2}{\footnotesize$-2$}
\psfrag{-4}{\footnotesize $-4$}
\psfrag{2}{\footnotesize $2$}
\psfrag{4}{\footnotesize 4}
} 
\parbox{13cm}{
\caption{Pairs of rays intersecting at angles $\phi=0,\pi/4, \pi/2, 3\pi/4$ get 
mapped by different conformal transformations to pairs of 
intersecting arcs, interpolating between the circle and a pair of antiparallel lines.
\label{fig:rays}}}
\end{center}
\end{figure}

In this picture the path is given by
\beq
\label{cusp-def}
x^1=s\cos\frac{\phi}{2}\,,
\qquad
x^2=|s|\sin\frac{\phi}{2}\,,
\qquad
\Theta^1=\cos\frac{\theta}{2}\,,
\qquad
\Theta^2=\sign(s)\sin\frac{\theta}{2}\,.
\eeq
We can perform a conformal transformation which maps the point at infinity to finite distance, 
so the pair of rays get replaced by two arcs, intersecting at angle $\pi-\phi$, 
as in Figure~\ref{fig:rays}. We can take them 
to be arcs of circles of radius $r=1/(1-\sin(\phi/2))$ centered at $\pm(r-1)$. These arcs 
pass through the points $\pm1$. The distance between the two intersection points is 
$2r\cos(\phi/2)$ and diverges for $\phi\to\pi$ like $8/(\pi-\phi)$. In this limit the conformal 
transformation of the cusp approximates a pair of antiparallel lines.

Cusped Wilson loops suffer from logarithmic divergences 
\cite{Brandt:1981kf,Brandt:1982gz}. 
This is exactly the same as the linear time divergence of \eqn{2-lines-vev}. 
The expectation value of the cusped loop is therefore 
\beq
\vev{W_\text{cusp}}\approx \exp\Big[{-}\log(R/\epsilon)\,V(\phi,\theta,\lambda)\Big]\,.
\label{cusp-vev}
\eeq
The cutoffs of the two calculations are related by $\log(R/\epsilon)\sim T$.

In the case of the straight line, the conformal transformation gives the circle. Their 
expectation value is not the same, and this can be attributed to the large 
conformal transformation relating them \cite{dg-mm}. The same is true for 
cusped loops, in the special cases when $\phi=\pm\theta$ \cite{dgrt-big} 
(see also \cite{bgps,young}). 
In these cases, the loop preserves some supercharges and is finite. Then 
this finite quantity is tractable, and indeed the only interesting quantity to calculate. 
For generic angles a divergence arises which completely masks this 
finite term.%
\footnote{Due to this fact, in the present circumstances the line and the circle 
are equivalent and we will not recover the matrix model describing the expectation 
value of the circle \cite{esz, dg-mm,pestun} in the $\theta=\phi=0$ limit.}
Still, one has to make sure that the same prescription and regularization 
is used for both calculations. This is particularly true in the Lorentzian case, 
where the conformal map eliminates more than one point from space.

In the limit that $\phi\to\pi$ there will be an extra pole in $V$, as the lines become 
coincident. The residue at this pole is the potential between a pair of 
antiparallel lines in flat space, as we discuss in Section~\ref{sec:anti}.

\section{Weak coupling}
\label{sec:pert}

Instead of calculating the correlator of two lines on $\bS^3\times\bR$ we work with the 
cusp in $\bR^4$. The logarithmic divergence arising from such singular points in the 
loop were discussed extensively (see {\em e.g.}, 
\cite{Polyakov:1980ca,Brandt:1981kf,Brandt:1982gz,kr-wl, Korchemsky:1988si, Collins:1989bt}).

Allowing for the extra angle $\theta$ in $\cN=4$ SYM, the calculation of the potential 
$V^{(1)}$ at one--loop order was done in \cite{dgo}. The result is
\beq
V^{(1)}(\phi,\theta)=-2\,\frac{\cos\theta-\cos\phi}{\sin\phi}\,\phi\,.
\label{1-loop}
\eeq

The extension to 2--loops in the case when $\theta=0$ was done in \cite{mos} 
(see also \cite{kr-wl}). The resulting expressions were written in \cite{mos} in 
integral form, and in Appendix~\ref{app:integrals} we extend the expressions 
to $\theta\neq0$ and compute the integrals in closed form. 
The result can be written as a sum of the contribution of ladder graphs 
(after subtracting the exponentiation of the $O(\lambda)$ term), and the 
interacting graphs
\bal
V^{(2)}(\phi,\theta)
&=V^{(2)}_\text{lad}(\phi,\theta)+V^{(2)}_\text{int}(\phi,\theta)
\\
V^{(2)}_\text{lad}(\phi,\theta)
&=-4\,\frac{(\cos\theta-\cos\phi)^2}{\sin^2\phi}
\left[\Li_3\left(e^{2i\phi}\right)-\zeta(3)
-i\phi\left(\Li_2\left(e^{2i\phi}\right)+\frac{\pi^2}{6}\right)
+\frac{i}{3}\phi^3\right],
\\
V^{(2)}_\text{int}(\phi,\theta)
&=\frac{4}{3}\,\frac{\cos\theta-\cos\phi}{\sin\phi}
\,(\pi-\phi)(\pi+\phi)\phi\,.
\label{2-loop}
\eal

To check these analytic expressions, one can focus on the BPS case \cite{zarembo}, 
when $\phi=\pm\theta$ and indeed $V^{(1)}=V^{(2)}=0$ as expected. 
As another test, for large imaginary angle the leading behavior is 
\bal
V^{(1)}(iu,\theta)&=2u+O(e^{-u})\,,\\
V^{(2)}(iu,\theta)&=-\frac{2\pi^2}{3}u-4\,\zeta(3)+O(u^{-1})\,.
\eal
together we find
\beq
V(iu,\theta)=\left(\frac{\lambda}{8\pi^2}-\frac{\lambda^2}{384\pi^2}+O(\lambda^3)\right)u+O(u^0)\,,
\eeq
and the prefactor of the linear term indeed matches a quarter of the perturbative 
expansion of the cusp anomalous dimension, $\gamma_\text{cusp}$ 
\cite{Kotikov:2003fb}.

Note that both interacting and ladder graphs at this order have uniform transcendentality 
three (when $e^{2i\phi}$ is considered rational). 
It is rather interesting that the complicated interacting graphs give a result which is much 
simpler than the 2--loop ladder graph and does not involve polylogarithmic functions. 
Indeed it is proportional to the 1--loop result
\beq
V^{(2)}_\text{int}(\phi,\theta)
=-\frac{2}{3}(\pi^2-\phi^2)V^{(1)}(\phi,\theta)\,.
\eeq
The fact that the prefactor $\cos\theta-\cos\phi$ (where all the $\theta$ dependence lies) 
is the same is obvious and can be seen also before integration. It is much more 
intriguing that the ratio of the final result of integration and $V^{(1)}$ is a polynomial 
in $\phi$.

In particular, in \cite{mos} an integral equation was written whose solution gives the 
contribution of ladder graphs to all orders in perturbation theory. If the interacting 
graphs are all related to lower order ladder graphs by some simple relations, it may 
be possible to find a full expression for $V(\phi,\theta,\lambda)$ for all values of the coupling.

\begin{figure}[t]
\begin{center}
\epsfig{file= 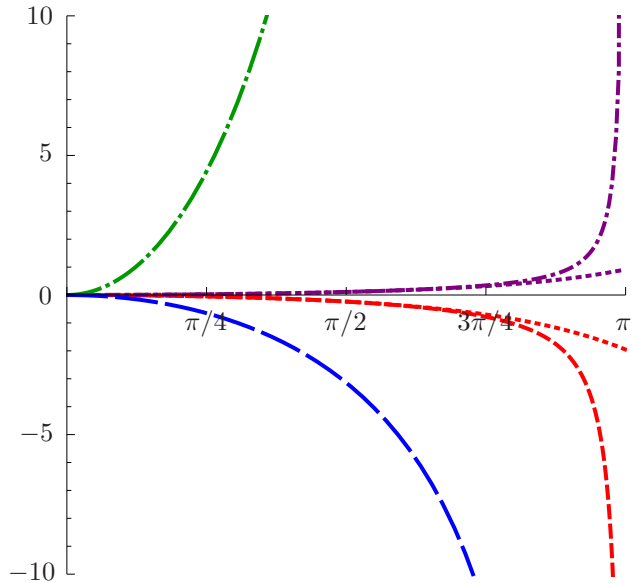,width=80mm
\psfrag{m10}{\footnotesize $\hskip-8pt-10$}
\psfrag{m5}{\footnotesize $\hskip-7pt -5$}
\psfrag{0}{\footnotesize $\raisebox{-1pt}{\hskip-5pt0}$}
\psfrag{5}{\footnotesize $\hskip-5pt5$}
\psfrag{10}{\footnotesize $\raisebox{-1pt}{\hskip-7pt10}$}
\psfrag{pi4}{\raisebox{-5pt}{\footnotesize $\hskip-4pt\pi/4$}}
\psfrag{pi2}{\raisebox{-5pt}{\footnotesize $\hskip-4pt\pi/2$}}
\psfrag{ppp4}{\raisebox{-5pt}{\footnotesize $\hskip-4pt3\pi/4$}}
\psfrag{p}{\raisebox{-5pt}{\footnotesize $\hskip-2pt\pi$}}
}
\parbox{13cm}{
\caption{Curves showing $V^{(1)}(\phi,0)$ (blue, wide dashes), 
$V^{(2)}(\phi,0)$ (green, dash-dot), 
$V^{(0)}_{AdS}(\phi,0)$ (red, short dashes)
And $V^{(1)}_{AdS}(\phi,0)$ (purple, short dash-dot). 
Note that all have a simple pole at $\phi=\pi$, with different residues.
The dotted lines show the perturbative expansions of 
$V^{(0)}_{AdS}$ and $V^{(1)}_{AdS}$ around $\phi=0$ to 
order $\phi^8$, \eqn{V0ads-expand}, \eqn{V1ads-expand}, which 
furnish a good approximation up to $\phi\sim2$.
\label{fig:curves}}}
\end{center}
\end{figure}

\section{Strong coupling}
\label{sec:ads}

In the strong coupling $AdS$ dual, Wilson loops are described by macroscopic 
strings \cite{maldacena-wl, rey-yee}. 
It is easy to write down the classical string solutions in $AdS_5\times\bS^5$ 
describing these Wilson loops. For the gauge theory on $\bS^3\times\bR$ 
it is appropriate to use global Lorentzian $AdS_5$ and take an ansatz which 
is time independent. For the cusp in $\bR^4$ it is more appropriate to consider 
the Euclidean Poincar\'e patch, in which case the ansatz posses a conformal 
symmetry. The solutions are clearly related by an isometry which is the bulk 
extension of the conformal transformation discussed in Section~\ref{sec:setup}.

These classical solutions were written down in the case of $\theta=0$ in 
\cite{dgo} and for $\theta\neq0$ in Appendix C.2 of \cite{dgrt-big}. As we review 
in Appendix~\ref{app:classical}, 
the classical solutions are expressed as functions of two parameters 
$q$ and $p$ \eqn{qp} (or in terms of $b$ and $k$ \eqn{b}, \eqn{xi}) 
and can be found for arbitrary values of $\phi$ and $\theta$, 
as the solutions of transcendental equations.

The quadratic fluctuation Lagrangian can be written for all values of the 
parameters, see Appendix~\ref{app:fluc}. In two limits the mass matrix 
diagonalizes, which are for $\theta=0$ (equivalently $q=0$), and for 
$\phi=0$ (the limit $p\propto q\to\infty$). In both these cases (studied in 
Appendices~\ref{app:theta=0} and~\ref{app:phi=0} respectively) 
all the quadratic fluctuation operators can be written in the form 
of single gap Lam\'e differential operators.%
\footnote{There are many generalizations to the Lam\'e operator which 
may allow to calculate the determinants away from these limits.}
The determinants for each of these operators can be calculated analytically, 
using the Gelfand-Yaglom method. The full determinant which includes the contribution 
of the trivial time direction is then expressed as a single integral, 
see equations \eqn{gammareg} and \eqn{gammaregbis}.

This is exactly the same form as was found for the case of the antiparallel lines 
in flat space in \cite{vali-lines} and it can be readily evaluated numerically 
for arbitrary values of $\phi$ and $\theta$, see the short daash-dot (purple) line in 
Figure~\ref{fig:curves}. In general we do not know how to calculate these 
integrals analytically, but we can evaluate them in a systematic expansion 
around $\theta=0$ and $\phi=0$.

Indeed, one motivation for studying this generalization of the antiparallel lines 
is that it would be easier to calculate it for $\theta\sim\phi\sim0$ than for 
$\phi=\pi$. We present the results of these expansions 
in Section~\ref{sec:expand} below.

\section{Antiparallel lines limit}
\label{sec:anti}

As mentioned in the introduction, part of the motivation for this project is as a stepping 
stone to understanding the potential between two antiparallel lines 
in $\cN=4$ SYM at all couplings. 
We introduced two deformation parameters, $\phi$ and $\theta$, and claimed that 
the original problem is recovered for $\phi\to\pi$ (and $\theta=0$).

To some degree this claim is obvious, one can look at figures~\ref{fig:hyperbolas} 
and~\ref{fig:rays}, and see that for $\phi\to\pi$ the curves approach antiparallel lines. 
Yet, this approach to the problem introduces a specific regularization prescription. 
We therefore examine here the $\phi\to\pi$ limit in detail and see how to recover the 
usual result for antiparallel lines from it.

\subsection{Weak coupling}
\label{sec:anti-weak}

Taking the $\phi\to\pi$ limit on the perturbative expressions \eqn{1-loop} and 
\eqn{2-loop} leads to a pole
\beq
V_{||}(\phi,\theta)\to-\frac{\lambda}{8\pi}\frac{1+\cos\theta}{\pi-\phi}
+\frac{\lambda^2}{32\pi^3}\frac{(1+\cos\theta)^2}{\pi-\phi}\log\frac{e}{2(\pi-\phi)}
+O(\lambda^3)\,.
\label{pert-par}
\eeq

The potential between antiparallel lines was calculated for $\theta=0$ to two--loop order 
in \cite{essz}. The result found there is
\beq
V_{||}(0)=-\frac{\lambda}{4\pi L}+\frac{\lambda^2}{8\pi^3 L}\log\frac{T}{L}+O(\lambda^3)\,.
\eeq
This behavior indeed matches the leading pole we find in \eqn{pert-par}, with the 
replacement $L\to\pi-\phi$. The extra logarithmic divergence at two--loop order 
breaks the scaling behavior expected for the Wilson loop. Such divergences were 
explained in \cite{adm} as arising from infrared 
effects, and get replaced by a logarithm of the coupling when including higher order 
soft--gluon graphs, see the more careful treatment in \cite{Pineda}.

\subsection{Strong coupling}
\label{sec:anti-ads}

We can look at the same limit at strong coupling. The classical solution is written down in 
Appendix~\ref{app:classical} and one can see that the relevant limit is 
\eqn{qp}, \eqn{xi}
\beq\label{antiparallel-limit}
p\to0\,,\qquad
\frac{q^2}{p}=\frac{1-2k^2}{k\sqrt{1-k^2}}\ \ \text{fixed.}\qquad 
0<k<1.
\eeq
$\phi$ \eqn{phi-relation} indeed approaches $\pi$
\beq
\label{piminusphi}
\pi-\phi=2\sqrt{p}\,\frac{\EE-(1-k^2) \KK}{\sqrt{k}(1-k^2)^{1/4}}\,,
\eeq
where $\KK=\KK(k^2)$ and $\EE=\EE(k^2)$ are complete elliptic integrals of the first 
and second kind, see Appendix~\ref{app:funcs} for their definitions and some of their properties.

The expression for $\theta$ \eqn{theta-relation} is
\beq
\theta=2 \sqrt{1-2k^2}\, \KK
\eeq
and the action \eqn{classical-action} is
\bal
\lim_{\phi\to\pi}V_{AdS}^{(0)}(\phi,\theta)
&=-\frac{4}{\sqrt{p}}\frac{\EE-(1-k^2) \KK}{\sqrt{k}(1-k^2)^{1/4}}
=-\frac{8}{\pi-\phi}\frac{\big(\EE-(1-k^2) \KK\big)^2}{k\sqrt{1-k^2}}
\label{ads-coinc}
\eal
This is exactly the result found in \cite{maldacena-wl} (with $k^2\to(1-l^2)/(2-l^2)$ and 
$\pi-\phi\to L$ and a factor of 2 in the definition of $g_{YM}^2$).

Specializing to the case of $\theta=0$ we need to set $k^2=1/2$ and 
the elliptic integrals can be expressed in terms of $\Gamma\left(\frac{1}{4}\right)^2$. 
We then find
\beq
V_{AdS}^{(0)}(\phi,0)=-\frac{16\pi^3}{(\pi-\phi)\Gamma\left(\frac{1}{4}\right)^4}\,,
\eeq
agreeing with the result of \cite{maldacena-wl,rey-yee} for the antiparallel 
lines in flat space with the replacement $\pi-\phi\to L$.

We can compare our calculation of the one--loop determinant in Appendix~\ref{app:theta=0} 
with that performed for the parallel lines in~\cite{chr,vali-lines}. These papers studied 
the $\theta=0$ case, and the fluctuation operators there are the same as those in 
\eqn{Z-q=0} for the case $k^2=1/2$. The two calculations do differ in the dependence 
on the time direction. After integrating over the world--sheet time direction 
we need to replace the cutoff on the world--sheet time $\cal T$ with the target space time 
$T$, which for $p\to0$ is \eqn{rescaling}
\beq
\frac{{\cal T}}{T}=\frac{1}{\sqrt{k}(1-k^2)^{1/4}\sqrt{p}}
=\frac{2}{\pi-\phi}\frac{\EE-(1-k^2)\KK}{k\sqrt{1-k^2}}\,.
\eeq
This factor gives the expected $1/(\pi-\phi)$ pole for generic $k$. In the case of $k^2=1/2$ the 
ratio is $\pi/(\pi-\phi)\KK(1/2)$, which upon the replacement $\pi-\phi\to L$, is indeed the 
rescaling done in~\cite{chr,vali-lines}. 

Thus, we see that the $\phi\to\pi$ limit does indeed reproduce the result for the antiparallel 
lines both at weak and strong coupling with the replacement of the pole $\pi-\phi\to L$.

\section{Near straight--line expansion}
\label{sec:expand}

The limit of $\phi\to\pi$ is interesting physically, capturing the potential between antiparallel 
lines in flat space. But it is really no simpler than the general case. The opposite limit, when 
$\phi\to0$ is indeed simple. In that case the cusp disappears and we are left with an 
infinite straight line in $\bR^4$, or a pair of antipodal lines on $\bS^3\times\bR$.

In this section we study the systematic expansion of $V(\phi,\theta,\lambda)$ in this limit. 
We then focus on specific terms in this expansion and try to learn how to evaluate them 
for all values of the coupling.

\subsection{Weak coupling}
\label{sec:pert-exp}

Expanding the results of the perturbative expressions \eqn{1-loop}, \eqn{2-loop} 
around $\phi=\theta=0$ we find
\bal
V^{(1)}(\phi,\theta)=&\,
\theta^2-\phi^2
-\frac{1}{12}(\theta^2-\phi^2)^2
+\frac{1}{360}(\theta^2-\phi^2)^2(\theta^2-3\phi^2)+O((\phi,\theta)^8)\,,\\
V^{(2)}(\phi,\theta)=&\,-\frac{2\pi^2}{3}(\theta^2-\phi^2)
+\frac{1}{18}(\pi^2(\theta^2-\phi^2)^2+6(\theta^2-\phi^2)(3\theta^2-\phi^2))
\\&-\frac{1}{540}(\pi^2(\theta^2-\phi^2)^2(\theta^2-3\phi^2)
+30(\theta^2-\phi^2)^2(3\theta^2-\phi^2))+O((\phi,\theta)^8)\,.
\eal
All the terms are proportional to $\theta^2-\phi^2$, and indeed we expect $V(\phi, \theta,\lambda)$ 
to vanish for $\theta=\pm\phi$, which are BPS configurations.

Note that the expansion of $V^{(2)}(\phi,\theta)$ has terms with $\pi^2$ and terms without. 
In fact, all the $\pi^2$ terms are proportional to $V^{(1)}(\phi,\theta)$
\beq
V^{(2)}(\phi,\theta)=-\frac{2\pi^2}{3}V^{(1)}(\phi,\theta)
+\frac{1}{3}(\theta^2-\phi^2)(3\theta^2-\phi^2)
-\frac{1}{18}(\theta^2-\phi^2)^2(3\theta^2-\phi^2)+\cdots
\eeq
As was pointed out in Section~\ref{sec:pert}, the contribution of the interacting 
two--loops graphs \eqn{2-loop} has a simple polynomial relation to the one--loop term
\beq
V^{(2)}_\text{int}(\phi,\theta)=-\frac{2}{3}(\pi^2-\phi^2)V^{(1)}(\phi,\theta)\,.
\eeq
All the terms with the extra $\pi^2$ come from this piece. The terms without 
the $\pi^2$ come from both the ladder part as well as from the $2\phi^2 V^{(1)}/3$ of 
the interacting graphs.

\subsection{Strong coupling}
\label{sec:ads-exp}

We can also expand the result of the string calculation around $\theta=\phi=0$. 
At the classical level we have an implicit relation between $V_{AdS}^{(0)}$ 
and $\phi$ and $\theta$, as all are functions of $p$ and $q$ \eqn{qp}. 
In the relevant limit the parameter $p$ is large and we can expand
\begin{align}
&\phi=\frac{\pi}{p}+\frac{\pi(3q^2-5)}{4p^3}+\frac{3\pi(15q^4-70q^2+63)}{64p^5}+\frac{5\pi(7q^2
(5q^4-45q^2+99)-429)}{256p^7}+O(p^{-9})
\nonumber\\*
&\theta=\frac{\pi q}{p}+\frac{\pi q(q^2-3)}{4p^3}+\frac{3\pi q(3q^4-30q^2+35)}{64p^5}+\frac{5\pi q
(5q^2(q^4-21q^2+63)-231)}{256p^7}+O(p^{-9})
\nonumber\\
&V^{(0)}_{AdS}=
\frac{\pi(q^2-1)}{p^2}+\frac{3\pi(q^4-6q^2+5)}{8p^4}
+\frac{15\pi(q^6-15q^4+35q^2-21)}{64p^6}
\nonumber\\*
&\hskip3.13cm
+\frac{35\pi(5q^8-140q^6+630q^4-924q^2+429)}{1024p^8}
+O(p^{-10})
\end{align}
These relations can now be inverted to yield
\bal
\label{V0ads-expand}
V_{AdS}^{(0)}(\phi,\theta)
=&\,
\frac{1}{\pi}(\theta^2-\phi^2)
-\frac{1}{8\pi^3}(\theta^2-\phi^2)\left(\theta^2-5\phi^2\right)
+\frac{1}{64\pi^5}(\theta^2-\phi^2)\left(\theta^4-14\theta
^2\phi^2+37\phi^4\right)\\&\,
-\frac{1}{2048\pi^7}(\theta^2-\phi^2)
\left(\theta^6-27\theta^4\phi^2+291\theta^2\phi^4-585\phi^6\right)
+O((\phi,\theta)^{10})\,.
\eal

At the one--loop order we did the calculation for the case of $\theta=0$ in 
Appendix~\ref{app:theta=0} and for $\phi=0$ in Appendix~\ref{app:phi=0}. The resulting 
expression in each case is an integral over the log of the ratio of many complicated elliptic 
functions. In the limit of small $\phi$ (for $\theta=0$) and small $\theta$ (for $\phi=0$) 
the modulus of the elliptic functions, $k$ vanishes. The small $k$ expansion of 
all the elliptic functions is a power series in regular hyperbolic functions. The integral 
over the log of the resulting expression can always be done result in the power 
series (\ref{V1pexp}), (\ref{V1pexpbis})
\bal
\label{V1ads-expand}
V_{AdS}^{(1)}(\phi,0)
=&\,
\frac{3}{2}\frac{\phi^2}{4\pi^2}
+\left(\frac{53}{8}-3\,\zeta(3)\right)\frac{\phi^4}{16\pi^4}
+\left(\frac{223}{8}-\frac{15}{2}\zeta(3)-\frac{15}{2}\zeta(5)\right)\frac{\phi^6}{64\pi^6}
\\&\,
+\left(\frac{14645}{128}-\frac{229}{8}\zeta(3)-\frac{55}{4}\zeta(5)
-\frac{315}{16}\zeta(7)\right)\frac{\phi^8}{256\pi^8}+O(\phi^{10})\,,
\\
V_{AdS}^{(1)}(0,\theta)
=&\,
{-}\frac{3}{2}\frac{\theta^2}{4\pi^2}
+\left(\frac{5}{8}-3\,\zeta(3)\right)\frac{\theta^4}{16\pi^4}
+\left(\frac{1}{8}+\frac{3}{2}\zeta(3)-\frac{15}{2}\zeta(5)\right)\frac{\theta^6}{64\pi^6}
\\&\,
+\left(-\frac{11}{128}-\frac{5}{8}\zeta(3)+\frac{25}{4}\zeta(5)
-\frac{315}{16}\zeta(7)\right)\frac{\theta^8}{256\pi^8}
+O(\theta^{10})\,.
\eal

The general mixed terms require calculating determinants of matrix valued differential operators, 
which we have not attempted to perform. Noticing that the terms in all the other expansions we 
are proportional to $(\theta^2-\phi^2)$ does allow to relates some of the mixed terms. So assuming 
the same is true for the one--loop determinant, the coefficient of the $\theta^2\phi^2$ term 
is $(-29+24\,\zeta(3))/(64\pi^4)$.

We plot in Figure~\ref{fig:curves} all the different $V(\phi,0)$ and also the curves the for expansions 
in \eqn{V0ads-expand} and \eqn{V1ads-expand} up to order $\phi^8$. The curves coincide 
quite well up to $\phi\sim2$.

\subsection{On the expansion coefficients}
\label{sec:expand-coefs}

Let us now focus on the first expansion coefficients around $\phi=\theta=0$. 
As we have seen, there are no linear terms and the quadratic terms are
\beq
\label{theta^2}
\frac{1}{2}\frac{\partial^2}{\partial\theta^2}V(\phi,\theta,\lambda)\Big|_{\phi=\theta=0}=
-\frac{1}{2}\frac{\partial^2}{\partial\phi^2}V(\phi,\theta,\lambda)\Big|_{\phi=\theta=0}=
\begin{cases}
\displaystyle
\frac{\lambda}{16\pi^2}-\frac{\lambda^2}{384\pi^2}+\cdots
\quad& \lambda\ll1\,,\\[4mm]
\displaystyle
\frac{\sqrt{\lambda}}{4\pi^2}-\frac{3}{8\pi^2}+\cdots
\quad& \lambda\gg1\,.
\end{cases}
\eeq

These expressions as well as all the higher derivative terms can be derived 
from studying the straight Wilson loop operator with operator insertions.
\beq
\frac{\partial^2}{\partial\theta^2}V(0,0)=
-\frac{1}{T}\frac{\partial^2}{\partial\theta^2}\log\vev{W}\approx
-\frac{1}{T}\frac{\partial^2}{\partial\theta^2}\vev{W}.
\eeq
The first identity is just the definition of $V$. The second is true since 
$\frac{\partial}{\partial\theta}\vev{W}=0$ due to flavor charge conservation, 
and because $\vev{W|_{\phi=\theta=0}}=1$.

The derivative with respect to $\phi$ is a modification of the shape of the curve 
which has to be written as an integral over a functional variation 
\beq
\frac{\partial}{\partial\phi}=\int_0^\infty ds
\left(x^1(s)\frac{\delta}{\delta x^2(s)}-x^2(s)\frac{\delta}{\delta x^1(s)}\right).
\eeq
The variation with respect to $\theta$ can also be written in this way, but 
since $\theta$ appears only in the explicit scalar coupling, the derivation is a 
little easier. After a trivial rotation we can write the straight ($\phi=0$) Wilson 
loop in the $x^1$ direction with arbitrary $\theta$ as ({\em c.f.}, \eqn{cusp-def})
\beq
W=\frac{1}{N}\Tr\cP\left[e^{\int_{-\infty}^0 (iA_1+\Phi_1)ds}
e^{\int_0^\infty (iA_1+\Phi_1\cos\theta+\Phi_2\sin\theta)ds}\right].
\eeq
The Wilson loop is taken such that it couples to the scalar 
$\Phi_1$ for all $s<0$ and to the linear combination 
$\Phi_1\cos\theta+\Phi_2\sin\theta$ for $s>0$. 
To reduce clutter we fixed the parameterization such that 
$|\dot x|=1$, so we can ignore the difference between $x^\mu(s_i)$ and $s_i$. 
Then we find
\bal
\frac{1}{2}\frac{\partial^2}{\partial\theta^2}V=
&\,-\frac{1}{\ln(R/\epsilon)}\frac{1}{2N}\int_0^\infty ds_1\int_0^\infty ds_2\,
\vev{\Tr\cP\left[\Phi_2(s_1)\Phi_2(s_2)\,e^{\int_{-\infty}^\infty (iA_1+\Phi_1)ds}\right]}
\\&\,
+\frac{1}{\ln(R/\epsilon)}\frac{1}{2N}\int_0^\infty ds_1\,
\vev{\Tr\cP\left[\Phi_1(s_1)\,e^{\int_{-\infty}^\infty (iA_1+\Phi_1)ds}\right]}.
\label{insertions}
\eal
As a functional derivation, the 
second line in \eqn{insertions} is a contact term, when both derivatives act at the 
same point. The path ordering symbol which is required for gauge invariance of 
a non-abelian Wilson loop also takes care of the scalar insertions; orders them 
with open Wilson lines connecting them and extending to infinity.

The analogous calculation for the variation with respect to $\phi$ will give 
after one differentiation an insertion of the field strength $sF_{21}(s)$. 
The contact term in the quadratic variation gives 
$s^2D_2F_{21}$. For simplicity we will concentrate on the scalar insertions, 
rather than the gauge field and field strength case. 

We need to calculate the two terms in \eqn{insertions}. The first one is very simple. 
At order $\lambda$ there is a single propagator contracting the two $\Phi_2$ fields 
giving
\bal
-\frac{1}{\ln(R/\epsilon)}\frac{\lambda}{8\pi^2}
\int_0^\infty ds_1\int_{s_1}^\infty ds_2\,
\frac{1}{(s_1-s_2)^2}\,.
\eal
Regularizing this graph in a natural way leads to a linear divergence, but 
no log divergent terms.

This is true also for all higher order contributions to this term, which are 
proportional to this one--loop expression. 
The reason is that if we write $\Phi_2=Z+\bar Z$ as a linear combination of two complex 
scalar fields (orthogonal to $\Phi_1$), then the insertion of two $Z$s or two $\bar Z$s 
vanishes due to charge conservation. The Wilson loop with one $Z$ and one $\bar Z$ 
insertion is BPS (unless the insertions are coincident), and receives no divergent quantum 
corrections \cite{dk-spinchain}.

The single insertion of $\Phi_1$ is more complicated. At order $\lambda$ we 
need to expand the Wilson loop to linear order and find
\bal
\frac{1}{2}\frac{\partial^2}{\partial\theta^2}V
=\frac{1}{\ln(R/\epsilon)}\frac{\lambda}{16\pi^2}\int_0^\infty ds_1\int_{-\infty}^\infty ds_2\,
\frac{1}{(s_1-s_2)^2}+O(\lambda^2)\,.
\eal
doing the integrals gives of course the same as we get by the expansion of 
$V^{(1)}$.
 
At order $\lambda^2$ there are several diagrams which contribute, but all of them 
include interactions. For the ladder graphs, at least one of the rangs will involve the 
Wilson loop alone, and it vanishes. This can be seen of course by the explicit 
expansion of $V^{(2)}_\text{int}$ and $V^{(2)}_\text{lad}$ in \eqn{2-loop}. This 
argument should apply also to higher order graphs, where only graphs with a single  
connected component attached to the Wilson loop contribute to this term.%
\footnote{This statement is true assuming the cancelation for the straight line 
does not require integration. Otherwise, there will be boundary terms in the 
disconnected graphs, which can be regarded as connected ones.}

To calculate this term directly in string theory will require to study a string 
world--sheet with the topology of a disc and one boundary vertex operator. 
It would be interesting to try to perform this calculation. Note that the insertion 
of $\Phi_1$ into the Wilson loop can be seen as a local change in the magnitude 
of the scalar coupling $\theta^I$ in \eqn{wl}. This takes the Wilson loop slightly 
away from the locally BPS condition \cite{dgo}, whose string theory interpretation 
was given in \cite{alday-mal2,pol-sul}.

In terms of the open spin--chain picture of deformations of Wilson loops 
\cite{dk-spinchain}, this is not a nice operator. An insertion into the 
straight Wilson loop can be assigned a conformal dimension which can be 
calculated by solving a spin--chain problem. For this spin--chain to be integrable 
(which was checked at order $\lambda$), the boundary conditions 
allow any of the scalar fields, but not $\Phi^1$ near the boundary. 
One cannot use in this way integrability to calculate this insertion of a single 
$\Phi^1$ into the straight line Wilson loop.

Still, we find that the contributions to the order $\theta^2$ term (and likewise 
$\phi^2$) should come only from graphs with one set of connected internal 
lines attached to the Wilson loop. The next order, like $\theta^4$, 
involve graphs with at most two disconnected internal  components, and so 
on. Note that we also found by explicit calculation that the connected (interacting) 
graphs at 2--loop order had a simpler functional form than the disconnected (ladder) 
ones, without polylogarithms. It would be interesting to see if this structure persists at 
higher orders in perturbation theory and whether it is possible to guess the answer 
for the most connected graphs at all loop order, and reproduce the strong 
coupling results in \eqn{theta^2}.

\section{Discussion}
\label{sec:disc}

We have studied a family of Wilson loop operators which continuously interpolates 
between the $1/2$ BPS line and the antiparallel lines. All these Wilson loops 
can be thought of as calculating a generalization of the quark--antiquark potential 
for the gauge theory on $\bS^3\times\bR$.

We have evaluated them in perturbation theory up to two loop order. In string theory 
we have the classical solutions for all these loops and the one--loop determinant 
for a pair of one parameter families. The determinant is given by an integral which 
can be evaluated numerically or, when expanded around the straight line configuration, 
analytically.

Therefore, when expanding around the straight line, for small $\phi$ and $\theta$ 
we have analytic results at both weak and strong coupling. It is tempting to try 
to guess interpolating functions satisfying the asymptotic behavior in 
\eqn{theta^2}, though we have refrained from doing so. We did argue, however, 
that this quantity receives contributions only from a subset of graphs in 
perturbation theory---the most connected graphs.

In the case of the circular Wilson loop, it turned out that in the Feynman gauge 
only ladder diagrams contribute and all interacting graphs combine to 
vanish \cite{esz, dg-mm,pestun}. Here we find instead an observable 
which gets contributions only from the most interacting graphs. Summing 
up ladder graphs is rather easy, but to our surprise, from the explicit calculation 
of the 2--loop graphs, we found that the result of these internally--connected graphs 
is simpler than the internally--disconnected one and does not involve polylogarithms. 
It would be very interesting to explore the 3--loop graphs and see whether 
a similar pattern persists and perhaps learn how to calculate the most 
connected graphs to all orders.

We have focused our discussion on loops in Euclidean space, but by 
analytic continuation $u=i\phi$ we can describe also loops in Minkowski space. 
In the limit of $u\to\infty$ the cusp becomes null and our calculations at weak coupling 
reproduce the known results for the cusp anomalous dimension. This quantity 
plays a crucial role in the calculation of scattering amplitudes in $\cN=4$ SYM. 
Since we study the system also away from this limit, our results could 
be useful for generalizations or regularizations of scattering amplitudes.

We have considered only the simplest generalization of the antiparallel lines 
geometry. There are many other deformations one can make and still 
retain the ability to find the minimal surfaces in $AdS$. Examples are 
helical curves in flat space or on $\bS^3\times\bR$ with extra rotations 
around $\bS^5$, as in \cite{Tseytlin:2002tr,df-int}. It may be possible to 
find other families of curves interpolating between the circle and the 
antiparallel lines using the techniques of \cite{Ishizeki:2011bf}.

\subsection*{Acknowledgements}

We are grateful to James Drummond, David Gross, Johannes Henn, 
Shoichi Kawamoto, Yuri Makeenko, 
Sanefumi Moriyama, Simon Scott, Domenico Seminara, 
for very useful discussions and to Juan Maldacena and Arkady Tseytlin for 
detailed comments on the manuscript. 
A preliminary version of these results was presented at the 
Nagoya University GCOE Spring School 2011 and N.D. is grateful for all the 
insightful questions and comments which helped improve this manuscript. 
We would like to thank Nordita (Stockholm) for the stimulating atmosphere 
during the ``Integrability in String and Gauge Theories; AdS/CFT Duality 
and its Applications'' workshop, where this project started. 
N.D. would also like to thank the KITP (Santa Barbara), POSTECH (Pohang), 
KIAS (Seoul), DESY (Hamburg), Humboldt University (Berlin), 
the Weizmann Institute (Rehovot), National Taiwan University (Taipei), 
Nagoya University and CERN
for their hospitality during the course of this work. V.F. thanks the 
Queen Mary University of London, the Galileo Galilei Institute in Florence during the 
``Large-$N$ Gauge Theories'' workshop and the Jagellonian University in Krakow for the kind hospitality and stimulating discussions had there. 
This research was supported in part by the National Science Foundation 
under Grant No. NSF PHY05-51164 and in part by INFN.
The work of N.D. is underwritten by an advanced fellowship of the 
Science \& Technology Facilities Council.

\appendix

\section{2--loop integrals}
\label{app:integrals}

The calculation of the 2--loop graphs for the Wilson loop with a cusp in the case when 
$\theta=0$ was done in \cite{mos}. 
It is trivial to generalize their results also for non-zero $\theta$. 
The resulting expression can be written as a sum of the contribution of ladder graphs 
(after subtracting the exponentiation of the $O(\lambda)$ term), which was already 
calculated in \cite{kr-wl} and the interacting graphs%
\footnote{\label{mos-footnote}To compare with \cite{mos}, one needs to replace $\theta\to\phi$ and 
in the numerators $1-\cos\theta\to\cos \theta-\cos \phi $. Also, we undid the 
expression of $Y$ in terms of Feynman--parameter integrals. Note that while 
most Feynman graphs in \cite{mos} were written for Wilson loops in the fundamental 
representation, the final results are quoted for the loop in the adjoint, which is double. 
Here all loops are in the fundamental.}
\bal
V^{(2)}(\phi,\theta)
&=V^{(2)}_\text{lad}(\phi,\theta)+V^{(2)}_\text{int}(\phi,\theta)
\\
V^{(2)}_\text{lad}(\phi,\theta)
&=-\frac{(\cos\theta-\cos\phi)^2}{\sin^2\phi}
\int_0^\infty \frac{dz}{z}\log\left(\frac{1+z\,e^{i\phi}}{1+z\,e^{-i\phi}}\right)
\log\left(\frac{z+e^{i\phi}}{z+e^{-i\phi}}\right)
\\
V^{(2)}_\text{int}(\phi,\theta)
&=4(\cos\theta-\cos\phi)
\int_0^1 dz \,Y(z^2,z^2+2z\cos\phi+1,1)\,.
\label{2-loop-int}
\eal
The integrand in the last expression is the ``scalar triangle graph'' --- the Feynman 
diagram arising at one--loop order from the cubic interaction between 
three scalars separated by distances given by the arguments
\beq
Y(x_{12}^2,x_{23}^2,x_{13}^2)=\frac{1}{\pi^2}
\int d^4w \frac{1}{|x_1-w|^2|x_2-w|^2|x_3-w|^2}\,,
\qquad
x_{ij}^2=|x_i-x_j|^2\,.
\eeq
This integral is known in closed form. For $x_{12}^2,x_{23}^2<x_{13}^2$ it is equal to \cite{davydychev}
\beq
\begin{aligned}
Y(x_{12}^2,x_{23}^2,x_{13}^2)=\frac{1}{x_{13}^2A}\bigg[
&\frac{\pi^2}{3}-2\Li_2\left(\frac{1+s-t-A}{2}\right)
-2\Li_2\left(\frac{1-s+t-A}{2}\right)
\\&
-\ln s\ln t
+2\ln\left(\frac{1+s-t-A}{2}\right)\ln\left(\frac{1-s+t-A}{2}\right)\bigg]\\
s=\frac{x_{12}^2}{x_{13}^2}\,,\qquad
t&=\frac{x_{23}^2}{x_{13}^2}\,,\qquad
A=\sqrt{(1-s-t)^2-4st}\,.
\end{aligned}
\eeq
This expression is valid for $s,t<1$ with the principle branch of the logarithms and 
dilogarithms. If $x_{12}^2$ is the largest, then the result is the same function divided by 
$s$ and the replacement $s\to1/s$ and $t\to t/s$. Likewise when $x_{23}^2$ is 
the largest.

In our case, if we take $\phi>2\pi/3$, then for $z\leq1$ the first two arguments of $Y$ in 
\eqn{2-loop-int} are less than unity and in that regime $Y$ evaluates to
\beq
\begin{aligned}
Y=-\frac{i}{z\sin\phi}\bigg(&\frac{\pi^2}{6}
-\Li_2\left(-ze^{i\phi}\right)
-\Li_2\left(1+ze^{-i\phi}\right)
-\log(z)\log\left(1+ze^{i\phi}\right)
\\&
+\log\left(-e^{i\phi}\right)\log\left(1+ze^{-i\phi}\right)
\bigg).
\end{aligned}
\eeq
The integration then gives
\beq
\int_0^1 dz\,Y=\frac{(\pi-\phi)(\pi+\phi)\phi}{3\sin\phi}\,.
\eeq
With the prefactor we find the final expression (valid by analytical continuation for 
all $0\leq\phi<\pi$) is
\beq
V^{(2)}_\text{int}(\phi,\theta)
=\frac{4}{3}\,\frac{\cos\theta-\cos\phi}{\sin\phi}(\pi-\phi)(\pi+\phi)\phi\,.
\eeq

The first integral in \eqn{2-loop-int} can also be done analytically. Again one should 
take care in choosing branch cuts for the logarithms, where the principle branch 
is for small $\phi$. The result is
\bal
V^{(2)}_\text{lad}(\phi,\theta)
=-4\frac{(\cos\theta-\cos\phi)^2}{\sin^2\phi}
\left[\Li_3\left(e^{2i\phi}\right)-\zeta(3)
-i\phi\left(\Li_2\left(e^{2i\phi}\right)+\frac{\pi^2}{6}\right)
+\frac{i}{3}\phi^3\right].
\eal

\section{Classical string solutions}
\label{app:classical}

We rederive here the classical string solutions dual to our Wilson loops with arbitrary 
$\theta$ and $\phi$ presented first in Appendix C.2 of \cite{dgrt-big}. 
The strong coupling dual of the gauge theory on $\bS^3\times\bR$ is string theory on 
global Lorentzian $AdS_5\times\bS^5$. The boundary conditions are within an 
$\bR\times\bS^1\times\bS^1$ on the boundary and therefore it suffices to consider an 
$AdS_3\times\bS^1$ subspace with metric (in units of $\alpha'=1$)
\beq
ds^2=\sqrt{\lambda}\left(-\cosh^2\rho\,dt^2+d\rho^2+\sinh^2\rho\,d\varphi^2+d\vartheta^2\right).
\label{metric1}
\eeq
As world--sheet coordinates we can take $t$ and $\varphi$ (with Lorentzian metric) 
and due to translation invariance the two other coordinates are functions of $\varphi$ 
alone
\beq
\rho=\rho(\varphi)\,,\qquad\vartheta=\vartheta(\varphi)\,.
\eeq
The coordinate $\varphi$ will vary in the domain $[\phi/2,\pi-\phi/2]$. At the 
two boundaries the coordinate $\rho$ should diverge, while for $\varphi=\pi/2$ it 
should take its minimal value. $\vartheta$ extends from $-\theta/2$ to $\theta/2$.

The Nambu-Goto action is
\beq
\cS_{NG}=\frac{\sqrt\lambda}{2\pi}\int dt\,d\varphi
\cosh\rho\sqrt{\sinh^2\rho+(\partial_\varphi\rho)^2+(\partial_\varphi\vartheta)^2}\,.
\label{NG}
\eeq
The time dependence is trivial and it is easy to find two conserved
quantities, the Hamiltonian (for $\partial_\varphi$ translations) 
and the canonical momentum conjugate to $\vartheta$
\beq
E=-\frac{\sinh^2\rho\cosh\rho}{\sqrt{\sinh^2\rho+(\partial_\varphi\rho)^2+(\partial_\varphi\vartheta)^2}}\,,
\qquad
J=\frac{\partial_\varphi\vartheta\cosh\rho}{\sqrt{\sinh^2\rho+(\partial_\varphi\rho)^2+(\partial_\varphi\vartheta)^2}}\,.
\label{EJ}
\eeq
The case of interest in \cite{dgrt-big} was when $E=\pm J$, which turns up to imply 
$\phi=\pm\theta$ and the string configuration preserves eight supercharges \cite{zarembo}. 
In these cases the solution is given in terms of trigonometric and hyperbolic functions. 
In the general case, when $\phi$ and $\theta$ are arbitrary, the solution is given in terms 
of elliptic integrals. See Appendix~\ref{app:funcs} for the definition and properties 
of some of these functions.

We denote
\beq
q=-\frac{J}{E}=\frac{\partial_\varphi\vartheta}{\sinh^2\rho}\,,
\qquad
p=\frac{1}{E}\,.
\label{qp}
\eeq
Using this we find the differential equation for $\rho$
\beq
(\partial_\varphi\rho)^2=p^2\cosh^2\rho\sinh^4\rho-q^2\sinh^4\rho-\sinh^2\rho\,,
\label{rho-eqn}
\eeq
This is an elliptic equation. To see that define
\beq
\label{b}
\xi=\frac{1}{b}\sqrt{\frac{b^4+p^2}{b^2+p^2\sinh^2\rho}}\,,\qquad
b^2=\frac{1}{2}\left(p^2-q^2+\sqrt{(p^2-q^2)^2+4p^2}\right).
\eeq
Then $\xi$ satisfies
\beq
(\partial_\varphi\xi)^2=\frac{b^2(b^4+p^2)}{p^4}
\left(1-\frac{b^4+p^2}{b^4\xi^2}\right)^2(1-\xi^2)(1-k^2\xi^2)\,,
\qquad
k^2=\frac{b^2(b^2-p^2)}{b^4+p^2}\,.
\label{xi}
\eeq
Therefore the relation between $\xi$ and $\varphi$ is given in terms of incomplete 
elliptic integrals of the first and third kind $F$ and $\Pi$ with argument 
$\arcsin\xi$ and modulus%
\footnote{Using Abramowitz \& Stegun/Mathematica notation.}
$k$
\beq\label{eqphixi}
\varphi=\frac{\phi}{2}+\frac{p^2}{b\sqrt{b^4+p^2}}
\left[\Pi\big({\textstyle\frac{b^4}{b^4+p^2}},\arcsin\xi|k^2\big)-F(\arcsin\xi|k^2)\right]
\eeq
The integration constant was chosen such that at the boundary, where 
$\rho\to\infty$ and $\xi=0$, the world--sheet coordinate takes its limiting value 
$\varphi\to\phi/2$. The maximum value of $\xi$ is $\xi=1$, which should 
happend for $\varphi=\pi/2$. Therefore $\phi$ is given in terms as the complete 
elliptic integrals by
\beq
\phi=\pi-2\frac{p^2}{b\sqrt{b^4+p^2}}
\left[\Pi\big({\textstyle\frac{b^4}{b^4+p^2}}|k^2\big)-\KK(k^2)\right]
\label{phi-relation}
\eeq
For $\pi/2<\varphi<\pi-\phi/2$ we need to analytically continue the solution. In that 
regime it is given by
\beq
\varphi=\pi-\frac{\phi}{2}-\frac{p^2}{b\sqrt{b^4+p^2}}
\left[\Pi\big({\textstyle\frac{b^4}{b^4+p^2}},\arcsin\xi|k^2\big)-F(\arcsin\xi|k^2)\right]
\eeq
Note that for notational brevity, we have chosen to express the solution in terms of 
$p$, $b$ and $k$, (and below also $q$), though there are algebraic relations among 
them \eqn{b}, \eqn{xi}, {\em i.e.},
\beq
\begin{gathered}
p^2=\frac{b^4(1-k^2)}{b^2+k^2}\,,\qquad
q^2=\frac{b^2(1-2k^2-k^2b^2)}{b^2+k^2}\,.
\end{gathered}
\eeq

Integrating $\vartheta$ gives a simple expression in terms of elliptic integrals of the first kind
\beq
\vartheta=\int d\varphi\,q\sinh^2\rho
=-\frac{\theta}{2}+\frac{b\,q}{\sqrt{b^4+p^2}}F(\arcsin\xi|k^2)\,.
\eeq
For $\xi=0$ it starts at $\vartheta=-\theta/2$ and for $\xi=1$ it reaches the midpoint 
$\vartheta=0$. Therefore
\beq
\theta=\frac{2b\,q}{\sqrt{b^4+p^2}}\KK(k^2)\,.
\label{theta-relation}
\eeq

Then we can calculate the classical action
\beq
\begin{aligned}
\cS_\text{cl}&=\frac{\sqrt\lambda}{2\pi}\int dt\,d\varphi\,
p\cosh^2\rho\sinh^2\rho
\\
&=\frac{T\sqrt\lambda}{2\pi}\frac{\sqrt{b^4+p^2}}{b\,p}
\left[-\frac{\sqrt{(1-\xi^2)(1-k^2\xi^2)}}{\xi}
+\frac{(b^2+1)p^2}{b^4+p^2} F(\arcsin\xi|k^2)-E(\arcsin\xi|k^2)\right]
\end{aligned}
\eeq
where $T$ is a cutoff on the $t$ integral and 
$E$ denotes an elliptic integral of the second kind. The
right hand side should be evaluated at the two boundaries where
$\xi=0$ as well as the two midpoints $\xi=1$. The result is
\beq
\cS_\text{cl}=\frac{T\sqrt\lambda}{2\pi}\frac{2\sqrt{b^4+p^2}}{b\,p}
\left[\frac{1}{\xi_0}+\frac{(b^2+1)p^2}{b^4+p^2} \KK(k^2)-\EE(k^2)\right].
\label{classical-action}
\eeq
Here $\xi_0$ is a cutoff at small $\xi$, so the first term is equal to
\beq
\frac{T\sqrt\lambda}{2\pi}\frac{2\sqrt{b^4+p^2}}{bp\xi_0}
=\frac{2T\sqrt\lambda}{2\pi}\sinh\rho_0\,,
\eeq
where $\rho_0$ is a cutoff on $\rho$, and this is the standard linear divergence 
for two lines along the boundary. This divergence is canceled as usual by
a boundary term leaving the elliptic integrals in \eqn{classical-action}.

There are two interesting limiting cases of these solutions. The first is $q=0$, 
where $\theta=0$ and the string is entirely within $AdS_3$. The quantum 
fluctuations of this configuration are studied in detail in Appendix~\ref{app:theta=0}. 
The second is for $p\to\infty$ keeping $k$ finite (therefore $q$ and $b$ also 
diverge), where $\phi=0$. Now the classical solution is entirely within an 
$AdS_2\times\bS^1$ and its quantum fluctuations are studied in 
Appendix~\ref{app:phi=0}.

Instead of the antiparallel lines on $\bS^3\times\bR$ we can consider the 
cusp in $\bR^4$. In that case we use the Poincar\'e patch metric%
\footnote{With $\sinh u=r/z$ and $\tanh \tilde t=(r^2+z^2-1)/(r^2+z^2+1)$, where 
$\tilde t$ is the Wick rotation of $t$ in \eqn{metric1}.}
\beq
ds^2=\frac{\sqrt\lambda}{z^2}\left(dz^2+dr^2+r^2d\varphi^2\right)
+\sqrt\lambda\, d\vartheta^2\,.
\eeq
The cusp is located at the origin $r=0$ and is invariant under rescaling of $r$. 
This symmetry is then extended to the string world--sheet, where the $z$ coordinate 
will have a linear dependence on $r$. As (Euclidean) world--sheet coordinates we 
take $r$ and $\varphi$. The ansatz for the other coordinates is
\beq
z=r\,v(\varphi)\,,\qquad
\vartheta=\vartheta(\varphi)\,.
\eeq

The Nambu-Goto action is
\beq
\cS_{NG}=\frac{\sqrt\lambda}{2\pi}\int dr\,d\varphi
\frac{1}{r\,v^2}\sqrt{(\partial_\varphi v)^2+(1+v^2)(1+v^2(\partial_\varphi\vartheta)^2)}\,.
\eeq
The $r$ dependence is trivial and it is easy to integrate this system. Indeed, 
it is exactly the same action as \eqn{NG} with the identification
\beq
v=\frac{1}{\sinh\rho}\,,
\qquad
r=\exp t\,.
\eeq

\subsection{Conformal coordinates}

The induced metric on the string world--sheet is
\beq
\begin{aligned}
ds^2_\text{ind}
&=\sqrt{\lambda}\Big[{-}\cosh^2\rho\,dt^2
+\left((\partial_\varphi\rho)^2+\sinh^2\rho+(\partial_\varphi\vartheta)^2\right)d\varphi^2\Big]
\\&=\sqrt\lambda\cosh^2\rho\Big[{-}dt^2
+p^2\sinh^4\rho\,d\varphi^2\Big],
\end{aligned}
\eeq
where in the last line we used the equations of motion \eqn{qp} and \eqn{rho-eqn}.

Following \cite{dgt, chr, vali-lines}, we rewrite the metric in conformally flat form and use 
the resulting elliptic coordinates to evaluate the one--loop determinant in 
the following appendices. We note that 
$\sinh^2\rho\,d\varphi=d\vartheta/q$ and the most convenient choice of world sheet 
coordinates is then related to $\vartheta$ and $t$ by rescaling
\beq\label{rescaling}
\sigma=\frac{\sqrt{b^4+p^2}}{b\,q}\,\vartheta=F(\arcsin\xi|k^2)-\KK(k^2)\,,
\qquad
\tau=\frac{\sqrt{b^4+p^2}}{b\,p}\,t\,.
\eeq
The range of the world--sheet coordinates is (repressing the argument $k^2$)
\beq\label{range}
-\KK<\sigma<\KK\,,\qquad-\infty<\tau<\infty\,.
\eeq
The relation between $\sigma$ and $\xi$ can be inverted in terms of a Jacobi sn 
function with modulus $k$ (see Appendix~\ref{app:funcs})
\beq\label{rhocn}
\xi=\sn(\sigma+\KK)=\frac{\cn(\sigma)}{\dn(\sigma)}\,,
\qquad
\cosh^2\rho=\frac{b^4+p^2}{p^2b^2}\frac{1-k^2\xi^2}{\xi^2}
=\frac{1+b^2}{b^2\cn^2(\sigma)}
\eeq

The equations of motions \eqn{rho-eqn} are then equivalently written for $\rho(\sigma)$ as
(the prime is $'=\partial_\sigma$)
\beq\label{eqrhos}
\rho'^2=\frac{ (b^2\sinh^2\rho - 1) (b^2 + 
p^2\sinh^2\rho)}{(b^4+p^2)\sinh^2\rho}
\eeq
or, in terms of $\xi(\sigma)$, as
\beq
\xi'^2=1-k^2\xi^4-(1+k^2)\,\xi^2\,.
\eeq
It is also useful to write, from (\ref{qp}) and (\ref{eqphixi}), the 
equations of motion for $\vartheta(\sigma)$ and $\varphi(\sigma)$ 
\beq\label{eqthphs}
\vartheta'^2=\frac{p^2\,(b^2+1) -b^4}{b^4+p^2}\,,
\qquad
\varphi'^2=\frac{b^2}{(b^4+p^2)\sinh^4\rho}
\eeq

The induced metric is therefore
\beq
ds^2_\text{ind}
=\sqrt{\lambda}\,\frac{1-k^2}{\cn^2(\sigma)}\big[{-}d\tau^2+d\sigma^2\big].
\label{ind-metric}
\eeq

The 2--dimensional scalar curvature reads
\beq
R^{(2)}=-2\left(1+ \frac{k^2(1 + b^2)^2 }{b^4 (1 - k^2)\cosh^4\rho}\right)
=-2\left(1+\frac{k^2}{1-k^2}\cn^4(\sigma)\right).
\label{curvature}
\eeq

The coordinate $\varphi$ can be also expressed in terms of $\sigma$ as%
\beq
\label{changecoo}
\varphi=\frac{\pi}{2}+\frac{p^2}{b\sqrt{b^4+p^2}}\left(
\sigma-\Pi\big({\textstyle\frac{b^4}{b^4+p^2}},\am(\sigma+\KK)|k^2\big)
+\Pi\big({\textstyle\frac{b^4}{b^4+p^2}}|k^2\big)\right),
\eeq
where $\am(x)$ is the Jacobi amplitude. In particular, the initial value is
\beq
\frac{\phi}{2}=\frac{\pi}{2}-\frac{p^2}{b\sqrt{b^4+p^2}}
\left(\KK-\Pi\big({\textstyle\frac{b^4}{b^4+p^2}}|k^2\big)\right).
\eeq

\section{Fluctuation Lagrangean}
\label{app:fluc}

We would like to calculate now the fluctuation determinant about the family of classical 
solutions presented in Appendix~\ref{app:classical}. For the 
metric of the full $AdS_5\times\bS^5$ space we take
\bal
\label{adsmetric}
ds^2=&\cosh^2\rho\,dt^2+d\rho^2+\sinh^2\rho\left(
dx_1^2+\cos^2x_1(dx_2^2+\cos^2x_2\,d\varphi^2)\right)\\
&+dx_3^2+\cos^2x_3\Big(
dx_4^2+\cos^2x_4\big(dx_5^2+\cos^2x_5(dx_6^2+\cos^2x_6\,d\vartheta^2)\big)\Big).
\eal
We rescaled the metric by $1/\sqrt\lambda$, since it simplifies many of the expressions 
to follow and the one--loop determinant is anyhow independent of $\lambda$. 
We also Wick-rotated the metric to Euclidean signature. Both the rescaling and Wick 
rotation should be done to the induced metric \eqn{ind-metric} as well. 
The $AdS_3\times\bS^1$ subspace where the solution is located 
\eqn{metric1} is gotten by setting $x_i=0$.

\subsection{Bosons}

To evaluate the one--loop correction to the classical solution, we follow \cite{fgt,dgt} and 
expand the Nambu-Goto action to quadratic order in fluctuations near the classical background. 
We use the static gauge, where we set the fluctuations along the world--sheet directions 
to zero. Therefore, one should not consider fluctuations of $t$. The fluctuations of the 
other fields are
\beq
\rho=\rho(\sigma)+\delta\rho\,,\qquad
\varphi=\varphi(\sigma)+\delta\varphi\,,\qquad
\vartheta=\vartheta(\sigma)+\delta\vartheta\,,\qquad
x_i\,,\quad i=1,\cdots, 6\,.
\eeq
We still need to project out one direction parallel to the world--sheet in the $\sigma$ direction. 
It is possible to freeze, say, $\delta\varphi=0$, but it is more natural to chose two linear combinations 
of $\delta\rho$, $\delta\varphi$ and $\delta\theta$ which are normal to the world--sheet and freeze 
the third direction, which is tangential.

For the normal directions we take
\beq
\zeta_7=\frac{\vartheta'(\sinh^2\rho\,\varphi'\,\delta\varphi+\rho'\,\delta\rho)
-(\rho'^2+\sinh^2\rho\,\varphi'^2)\delta\vartheta}
{\sqrt{(\rho'^2+\sinh^2\rho\,\varphi'^2)(\vartheta'^2+\rho'^2+\sinh^2\rho\,\varphi'^2)}}\,,
\qquad
\zeta_8=
\frac{\sinh\rho\,(\rho'\,\delta\varphi-\varphi'\,\delta\rho)}
{\sqrt{\rho'^2+\sinh^2\rho\,\varphi'^2}}\,,
\eeq
where $\rho$, $\rho'$, $\varphi'$ and $\vartheta'$ are evaluated for the classical solution 
in (\ref{eqrhos}) and (\ref{eqthphs}).

Note that we chose $\zeta_7$ and $\zeta_8$ such that they are unit normalized, meaning that they 
will have canonical kinetic terms. For the fields $x_i$, two of them have to be rescaled by 
the vielbein to achieve the same
\beq
\zeta_i=x_i \sinh\rho\,,\quad i=1,2\,,\qquad\qquad
\zeta_s=x_s\,,\quad s=3,4,5,6\,,\qquad
\eeq

The resulting action takes the form
\beq
\label{bosfluct}
\cL_{B}=\frac{1}{2}\sqrt{g}\Big[g^{ab}\,\partial_a \zeta_P\,\partial_b \zeta_P
+A(\zeta_8\partial_\sigma\zeta_7-\zeta_7\partial_\sigma\zeta_8)
+M_{PQ}\zeta_P\zeta_Q\Big],
\qquad
P,Q=1,\cdots,8
\eeq
with
\begin{align}
A&=\frac{2\sqrt{b^4+p^2}\sqrt{-b^4+b^2p^2+p^2}}{p\left(b^4+b^2p^2\sinh^2\rho-p^2\right)\cosh^2\rho}
\nonumber\\*
M_{77}&=\frac{b^4-b^2p^2-p^2}{b^2p^2\cosh^2\rho}
-\frac{2(b^2+1)(b^2-p^2)}{b^2p^2\cosh^4\rho}
+b^2\frac{b^4+2b^2p^2\sinh^2\rho+b^2p^2-p^2}
{\cosh^2\rho\left(b^4+b^2p^2\sinh^2\rho-p^2\right)^2}
\,,
\nonumber\\
M_{78}&=M_{87}=\frac{2\sqrt{-b^4+b^2p^2+p^2}\sqrt{b^2\sinh^2\rho-1}\sqrt{b^2+p^2\sinh^2\rho}}
{p(b^4+b^2p^2\sinh^2\rho-p^2)\cosh^3\rho}\,,
\label{bosonmasses}
\\
M_{88}&=\frac{b^4-b^2p^2-p^2}{b^2p^2\cosh^2\rho}+2
-\frac{3b^2}{\cosh^2\rho(b^4+b^2p^2\sinh^2\rho-p^2)}
+\frac{b^4p^2}{\left(b^4+b^2p^2\sinh^2\rho-p^2\right)^2}\,,
\nonumber\\
M_{ii}&=\frac{b^4-b^2p^2-p^2}{b^2p^2\cosh^2\rho}+2\,,
\qquad\qquad\ i=1,2\,,
\nonumber\\*
M_{ss}&=\frac{b^4-b^2p^2-p^2}{b^2p^2\cosh^2\rho}\,,
\qquad\qquad\qquad s=3,4,5,6\,,
\nonumber
\end{align}
and all other entries of $M_{PQ}$ vanishing.

The first order terms can be eliminated by a rotation
\beq
\begin{pmatrix}\zeta_7\\\zeta_8\end{pmatrix}
\to\begin{pmatrix}\cos\alpha&\sin\alpha\\-\sin\alpha&\cos\alpha\end{pmatrix}
\begin{pmatrix}\zeta_7\\\zeta_8\end{pmatrix},
\eeq
where $\alpha(\sigma)$ solves the equation $\alpha'=-\frac{1}{2}\sqrt{g}A$
\beq
\alpha=\frac{b^2\sqrt{p^2+b^2p^2-b^4}}{p\sqrt{b^4+p^2}}
\left(\sigma+\KK-\sqrt{1-k^2}\,\Pi\big({\textstyle\frac{p^2}{b^4+p^2}},\am(\sigma+\KK)|k^2\big)\right).
\eeq
This also shifts the masses ($M_{PQ}\to \widetilde{M}_{PQ}$) by
\bal\label{bosonshift}
M_{77}\to&
\widetilde{M}_{77}=\frac{1}{2}\left(M_{77}+M_{88}+(M_{77}-M_{88})\cos2\alpha-M_{78}\sin2\alpha-\frac{\sqrt{g}}{2}A^2\right),
\\
M_{88}\to&\widetilde{M}_{88}=
\frac{1}{2}\left(M_{77}+M_{88}-(M_{77}-M_{88})\cos2\alpha+M_{78}\sin2\alpha-\frac{\sqrt{g}}{2}A^2\right),
\\
M_{78}\to&\widetilde{M}_{78}=
\frac{1}{2}\big(M_{78}\cos2\alpha+(M_{77}-M_{88})\sin2\alpha\big).
\eal

The two limiting cases where the classical solutions have either $\theta=0$ or $\phi=0$ 
lead to much simpler fluctuation operators, in particular the mass matrix is diagonal. 
We study them in detail in Appendices~\ref{app:theta=0} and~\ref{app:phi=0}.

\subsection{Fermions}

The derivation of the quadratic fermionic action is similar to the one in~\cite{dgt}, 
starting with the kinetic operator for Green-Schwarz fermions, as it is before gauge fixing 
\beq\label{fermlagr}
{\cal L}^\text{IIB}_{2F}=-i\Big(\sqrt{g}\,g^{ij}\delta_{IJ}-\epsilon^{ij}s_{IJ}\Big)\bar\psi^I\gamma_iD_j\psi^J
\eeq
where $s_{IJ}$ is defined by $s_{11}=-s_{22}=1$, $s_{12}=s_{21}=0$.
The $\gamma_i$ are the pullbacks to the worldsheet of the 10d gamma matrices 
\beq
\gamma_i=\Gamma_a\,e^a_\mu\,\partial_i X^\mu_\text{classical}
\eeq 
where $e^a_\mu$ are the vielbein for the metric (\ref{adsmetric}). The covariant
derivative $D_i$ is the projection of the 10-d derivative and has
the following explicit form~\cite{Metsaev:1998it} 
\beq\label{covdev}
D_i\psi^I=\Big(\delta^{IJ}{\cal D}_i-\frac{i}{2}\,\epsilon^{IJ}\,\Gamma_*\,\gamma_i\Big)\psi^I\,,
\qquad
{\cal D}_i=\partial_i+\frac{1}{4}\partial_i x^\mu\,\Omega_\mu^{ab}\Gamma_{ab}
\eeq
where the ``mass term'' originates from the coupling to the Ramond-Ramond 
field strength and can be written as  
\beq
\Gamma_*=i\,\Gamma_{a_0a_1a_2a_3a_4}\,.
\eeq
where $a_0,a_1,a_2,a_3,a_4$ are the tangent space indices of $AdS_5$.

To simplify the derivation we choose an auspicious frame%
\footnote{In \cite{dgt} and subsequent work a simple frame that simplifies the 
action was chosen only a-posteriori.}
which is
\bal
e^0&=\cosh\rho\,dt\,,
&\quad
e^9&=\frac{\rho'd\rho+\sinh^2\rho\,\varphi'd\varphi+\vartheta'd\vartheta}
{\sqrt{\rho'^2+\sinh^2\rho\,\varphi'^2+\vartheta'^2}}
\\
e^7&=\frac{\sinh\rho(\varphi'\,d\rho-\rho'\,d\varphi)}{\sqrt{\rho'^2+\sinh^2\rho\,\varphi'^2}}
&\quad
e^8&=\frac{\vartheta'(\rho'\,d\rho+\sinh^2\rho\,\varphi'\,d\varphi)
-(\rho'^2+\sinh^2\rho\,\varphi'^2)d\vartheta}
{\sqrt{(\rho'^2+\sinh^2\rho\, \varphi'^2)(\vartheta'^2+\rho'^2+\sinh^2\rho\,\varphi'^2)}}
\label{viel}
\eal
and the obvious (diagonal) choice for $e^1,\cdots,e^6$.

With this choice of frame it is automatic that $\gamma_i$ behave like 2d gamma matrices, {\em i.e.},
\beq\label{2dgamma}
\gamma_\tau
=\frac{\sqrt{1-k^2}}{\cn(\sigma)}\Gamma_0\,,
\qquad
\gamma_\sigma= \frac{\sqrt{1-k^2}}{\cn(\sigma)}\Gamma_9\,,
\qquad
\{\gamma_i,\gamma_j\}=g_{ij}\,,
\eeq
where we recall (\ref{rhocn}) that $\cn(\sigma|k^2)=\sqrt{1+b^2}/(b\cosh\rho)$.

The covariant derivative $D_j$ in \eqn{fermlagr} has the pullback of the bulk spin--connection 
to the world--sheet. To evaluate it we need to differentiate the expressions in \eqn{viel}, which 
we view as functions of $\rho$. 
To simplify the resulting expression it is useful to perform a further rotation
\beq
\begin{pmatrix}e^7\\e^8\end{pmatrix}
\to\begin{pmatrix}\cos\beta&\sin\beta\\-\sin\beta&\cos\beta\end{pmatrix}
\begin{pmatrix}e^7\\e^8\end{pmatrix},
\eeq
where (the constant term is for future convenience)
\beq
\label{beta}
\beta=
\frac{\pi}{2}+
\frac{b^4p}{b^2\sqrt{b^4+p^2}\sqrt{-b^4+b^2p^2+p^2}}
\textstyle
\left(\sigma-\frac{(b^2+1)p^2}{b^4}\,\Pi\big(\frac{b^4-b^2 p^2-p^2}{b^4},\am(\sigma)|k^2\big)\right)
\eeq
solves the equation
\beq
\partial_\rho\beta
=-\frac{b^2p\sqrt{-b^4+b^2p^2+p^2}\sinh\rho}
{\sqrt{b^2\sinh^2\rho-1}\sqrt{b^2+p^2\sinh^2\rho}\,(b^4+b^2p^2\sinh^2\rho-p^2)}\,.
\eeq

With this choice, by explicit computation, we find that the pullback of the covariant 
derivative to the world--sheet ${\cal D}_i$ is related to the 2d covariant derivative 
$\hat\nabla_i$ by
\bal
{\cal D}_\tau&=\partial_\tau+\frac{1}{4}\partial_\tau x^\mu\,\Omega_\mu^{ab}\Gamma_{ab}
=\hat\nabla_\tau+\Gamma_0\Gamma_\perp\,,
\\\label{covtautheta}
{\cal D}_\sigma&=\partial_\sigma+\frac{1}{4}\partial_\sigma x^\mu\,\Omega_\mu^{ab}\Gamma_{ab}
=\hat\nabla_\sigma-\Gamma_9\Gamma_\perp\,.
\eal
They are equal apart for shifts by $\Gamma_0\Gamma_\perp$ and $-\Gamma_9\Gamma_\perp$ with
\beq
\Gamma_\perp=\frac{b\,p\sinh\rho}{2\sqrt{b^4+p^2}}(e^7_\rho\Gamma_7+e^8_\rho\Gamma_8)\,.
\eeq
Similarly to what happens in~\cite{dgt, Frolov:2002av}, these extra terms cancel out in 
the fermionic action (\ref{fermlagr}) once contracted with $\gamma^i$ and with 
$\epsilon^{ij}\,\gamma_i$.

In (\ref{covtautheta}), the 2d covariant derivative
\beq
\hat\nabla_i=\partial_i+\frac{1}{4}\omega_i^{ab}\Gamma_{ab}
\eeq
has the world--sheet spin--connection
\beq
\omega_\tau^{ab}\Gamma_{ab}=
\frac{2\sqrt{b^2\sinh^2\rho-1}\sqrt{b^2+p^2\sinh^2\rho}}{\sqrt{b^4+p^2}\cosh\rho}
\,\Gamma_{09}\,,
\qquad
\omega_\sigma^{ab}\Gamma_{ab}=0\,,
\eeq
which can be nicely written in terms of Jacobi elliptic functions
\beq\label{spinconnection}
\omega_\tau^{ab}\Gamma_{ab}=
\frac{2\sn(\sigma)\dn(\sigma)}{\cn(\sigma)}\,\Gamma_{09}\,,
\qquad
\omega_\sigma^{ab}\Gamma_{ab}=0\,.
\eeq

In the new frame \ref{viel} the ``mass term'' in the covariant derivative becomes
\beq
\label{Gammastarbis}
\Gamma_*=\frac{i\,}{b\,p\cosh\rho}\Gamma_{012}
\left(\sqrt{b^4-p^2+p^2b^2\sinh^2\rho}\,(\Gamma_{79}\sin\beta+\Gamma_{89}\cos\beta)
-\sqrt{p^2+p^2b^2-b^4}\,\Gamma_{78}\right).
\eeq
Fixing then $\kappa$-symmetry 
\beq
\psi^1=\psi^2=\psi\,,
\eeq
one can check that the fermionic Lagrangean reads
\beq
\label{2d-fermi}
{\cal L}_{2F}=2\sqrt{g}\,\bar\psi\big(i\,\gamma^i\,\hat\nabla_i-M_F\big)\psi
\eeq
where
\beq\label{fermionmass}
M_F=i\frac{\sqrt{b^4-p^2+p^2b^2\sinh^2\rho}}{b\,p\cosh\rho}\,\Gamma_{12}(\Gamma_7\sin\beta+\Gamma_8\cos\beta)
\eeq

\subsection{Divergence cancellation}
\label{app:seeley}

Before attempting to evaluate the determinants explicitly, it is useful to check 
whether they are divergent or finite.

The quadratic and linear divergences cancel between bosons and fermions because
of the matching of the number of degrees of freedom. 
The coefficients of the logarithmic divergence for the system of bosonic and fermionic fields 
can be evaluated with the general formula for the relevant Seeley coefficient of the 
corresponding second order Laplace operators put in the standard form 
$\Delta=-g^{ab}\,D_a D_b+X$, which reads~\cite{Gilkey,Schwarz:1992te}
\beq\label{seeley}
b_2(\Delta)=\Tr\left(\frac{1}{6}\,\mathbf{I}\,R^{(2)}-X\right).
\eeq
In the case of the bosons, from the Lagrangean (\ref{bosfluct}) with (\ref{bosonshift}) one gets
\beq\label{b2bos}
b_2^\text{bos}
=\frac{1}{2}\left(8\times\frac{R^{(2)}}{6}-\Tr\widetilde{M}\right)
=\frac{2}{3}R^{(2)}-\frac{1}{2}\Tr\widetilde{M}\,,
\eeq
where 
\beq
\Tr\widetilde{M}
=M_{77}+M_{88}+2 M_{ii}+4\,M_{ss}-\frac{\sqrt{g}}{2}\,A^2\,,
\eeq

To put the fermionic Lagrangean (\ref{2d-fermi}) in the standard form one considers 
$D_F=i\gamma^i\,\hat\nabla_i-M_F$, and finds 
\beq
D_F^2 =-g^{ij}\,\hat \nabla_i\hat \nabla_j+\frac{R^{(2)}}{4}-i\,\gamma^\sigma\,\partial_\sigma M_F+M_F^2\,.
\eeq
Since $\Tr(\Gamma^9M_F)=0$, the logarithmic divergence of the operator $4\ln \det (D_F^2)$ is proportional to
(we regard $M_F$ as a $4\times4$ matrix)
\beq
b_2^\text{ferm}
=4\times\left(\frac{R^{(2)}}{6}-\frac{R^{(2)}}{4}\right)- \Tr(M_F^2)
= - \frac{R^{(2)}}{3}-\Tr(M_F^2)\,.
\eeq
Using the explicit expressions (\ref{bosonmasses}) and (\ref{fermionmass}) for the masses 
we find
\beq
-\Tr(M_F^2)+\Tr\widetilde{M}=R^{(2)}\,.
\eeq
Therefore the full divergent coefficient $b_2^\text{ferm}-b_2^\text{bos}$ is proportional to the 
world--sheet curvature $R^{(2)}$. The exact numerical coefficient does not matter, even 
though the integral over the curvature diverges \cite{bdfpt}. 
With proper regularization there should be boundary terms such that the total integral 
is proportional to the Euler character. In our case, when the world--sheet is the infinite 
strip, this vanishes.

This fact allows us to do the calculation in the static gauge ignoring ghosts (which are 
algebraic in this gauge, but should still contribute some curvature term to the Seeley 
coefficient \cite{dgt}). Likewise we do not have to worry about a subtlety in the measure 
of the Green-Schwarz fermions which has a similar effect 
\cite{Langouche:1987mx,Langouche:1987mw,Wiegmann:1989md,Lechner:1995yi}.

With the confidence that the proper determinants are finite, we will proceed to evaluate them 
in the next appendix in the special case when $\theta=0$ and in the following one for 
$\phi=0$. Despite the argument here, in the way the calculation is performed we ignore boundary 
terms which does lead to divergences. All these divergences are power--like, not logarithmic, 
so they can be easily removed without affecting the finite piece.

\section{One--loop determinant for $\theta=0$}
\label{app:theta=0}

In this appendix we focus on a special limit of the general fluctuation operator 
derived in Appendix~\ref{app:fluc}. We consider taking $q=0$ in the 
classical solution, which implies that the classical world--sheet is entirely inside 
an $AdS_3$ subspace of $AdS_5$.

In this limit we can express both $b$ and $p$ in terms of $k$ which takes values 
in the range $0\leq k<1/\sqrt2$ as 
$b^2=(1-2k^2)/k^2$ and $p^2=b^4/(1+b^2)$. The fluctuation mode 
$\zeta_7$ simplifies to $\zeta_7=\delta\theta$, 
and is similar to the other four fluctuation modes on $\bS^5$. 
In the quadratic Lagrangean \eqn{bosfluct} the mass parameters become
\beq
M_{11}=M_{22}=2\,,\qquad
M_{33}=\cdots=M_{77}=0\,,\qquad
M_{88}=R^{(2)}+4\,,
\eeq
where $R^{(2)}$ is the scalar curvature of the induced metric \eqn{curvature} and 
$A=0$.

In this limit, the angle $\beta=\pi/2$ \eqn{beta} and the mass term \eqn{fermionmass} 
in the fermionic fluctuation action \eqn{2d-fermi} gets significantly simplified
\beq\label{Mq0}
M_F=i\,\Gamma_{127}\,.
\eeq
Choosing a basis for the spinors where they are eigenstates of the five gamma matrices in 
the $AdS$ directions $\Gamma_{01279}$, this can be rewritten as a two-dimensional mass term
\beq
M_F=\frac{\epsilon^{ij}}{2\sqrt{g}}\,\gamma_i\gamma_j=\gamma_3.
\eeq
Namely, in this limit the fermionic partition function is
\beq\label{ZF}
Z_{F}=\det{}^4(i\gamma^i\,\hat\nabla_i-\gamma_3)\,.
\eeq

The resulting fluctuation problem gives the overall 1--loop correction to the partition function
\beq
\label{Z-q=0}
Z=\frac{\det^4(i\gamma^i\,\hat\nabla_i-\gamma_3)}
{\det(-\nabla^2+2)\,\det^{1/2}(-\nabla^2+R^{(2)}+4)\,\det^{5/2}(-\nabla^2)}\,,
\eeq
This is actually formally the same as the operators appearing in the case 
of the antiparallel lines~\cite{fgt,dgt}, and the different values of $\phi$ are 
distinguished by different world--sheet metrics.

\subsection{Lam\'e form of the fluctuation operators}
\label{app:q0-lame}

Since the world--sheet metric is conformal, after an overall rescaling%
\footnote{All the bosonic operators are rescaled by $\sqrt{g}$ and the fermionic one 
below (\ref{opferm}) by $g^{1/4}$. Such functional rescalings of operators can lead to 
extra logarithmic divergences. In this case, since the divergences of the original 
operators cancel and the rescaling are by these powers of the same function, 
no new divergences appear. See the discussion in Appendix A of~\cite{dgt}. 
The same applies to the operators (\ref{opbos0-2})-(\ref{opbos2-2}) 
and (\ref{fermopphi0})  of Appendix \ref{app:phi=0}.}
the derivatives in 
the bosonic differential operators appearing in (\ref{Z-q=0}) are just flat space operators. 
We can then Fourier transform the time direction $\partial_\tau\to i\omega$ and get the 
set of one--dimentional operators
\begin{align}
\label{opbos0}
\cO_0=\sqrt{g}\left(-\nabla^2\right)
&=-\partial_\sigma^2+\omega^2\,,\\
\label{opbos1}
\cO_1=\sqrt{g}\left(-\nabla^2+2\right)
&=-\partial_\sigma^2+\omega^2+\frac{2(1-k^2)}{\cn^2(\sigma)}\,,\\
\label{opbos2}
\cO_2=\sqrt{g}\left(-\nabla^2+R^{(2)}+4\right)
&=-\partial_\sigma^2+\omega^2+\frac{2(1-k^2)}{\cn^2(\sigma)}-2k^2\cn^2(\sigma)\,.
\end{align}
The first operator ${\cal O}_0=-\partial^2_\sigma+\omega^2$, is the free Laplacean. 
The other two operators can be transformed into single--gap Lam\'e operators 
(see {\em e.g.}, \cite{Whittaker}) 
with the typical form $-\partial_\sigma^2+2k^2\sn^2(\sigma|k^2)$. 
The rewriting involves modifying both the argument and the modulus of the elliptic 
functions, using identities which can be found for example in \cite{elliptic}. 

The second and third operator can be rewritten as
\begin{align}
\label{O1}
{\cal O}_1&=(1-k^2)\Big[{-}\partial_{\sigma_1}^2+\omega_1^2
+2k_1^2\sn^2(\sigma_1+i\KK_1'|k_1^2)\Big]\,,
&\quad k_1^2&=\frac{k^2}{k^2-1}\,,
\\
\label{O2}
{\cal O}_2&=(1-k^2)(1+k_1)^2\Big[{-}\partial_{\sigma_2}^2
+\omega_2^2+2k_2^2\sn^2(\sigma_2+i\KK_2'|k_2^2)\Big]\,,
&\quad k_2^2&=\frac{4k_1}{(1+k_1)^2}\,,
\end{align}
where the coordinates and the frequencies are rescaled as
\begin{align}
\label{sigma1}
\sigma_1&=\sqrt{1-k^2}\,\sigma+\KK_1, 
\qquad
&\omega_1^2&=\frac{\omega^2}{1-k^2}\,,\\
\label{sigma2}
\sigma_2&=(1+k_1)(\sqrt{1-k^2}\,\sigma+\KK_1), 
\qquad
&\omega^2_2&=\frac{\omega^2}{(1-k^2)(1+k_1)^2}-k_2^2\,,
\end{align}
and in all of
these formulae $\KK_i=\KK(k_i^2)$ and $\KK_i'=\KK(1-k_i^2)$.

Notice that the modulus $k_1^2$ is negative and $k_2^2$ complex. This however does not 
hinder the calculation of the determinants and the spectrum we find from it is positive definite.

Turning now to the fermions, the operator reads explicitly
\beq\label{opferm}
D_F=-i\gamma^i\,\hat\nabla_i+\gamma_3
=\frac{\cn(\sigma)}{\sqrt{1-k^2}}
\left[-i\left(\partial_\sigma+\frac{\sn(\sigma)\dn(\sigma)}{2\cn(\sigma)}\right)\tau_1
-\omega\,\tau_2+\frac{\sqrt{1-k^2}}{\cn(\sigma)}\,\tau_3\right],
\eeq
where $\tau_1,\tau_2,\tau_3$ are the three Pauli matrices.

As in~\cite{chr,vali-lines}, the fermionic differential operator 
$\frac{\sqrt{1-k^2}}{\cn(\sigma)}D_F$ 
can be further diagonalized after squaring it. Using 
$M =\textstyle{\frac{1}{\sqrt{2}}}\big( \begin{smallmatrix} 1&i\\ i&1 \end{smallmatrix}\big)$, 
one has
\beq\label{diag}
\left({\textstyle\frac{\sqrt{1-k^2}}{\cn(\sigma)}}D_F\right)^2
=\sqrt{\cn(\sigma)}\,M^\dagger\,{\diag}\{{\cal O}_+,\,{\cal O}_-\}\,M\frac{1}{\sqrt{\cn( \sigma)}}\,,
\eeq
where ${\cal O}_+$ and ${\cal O}_+$ are
\beq\label{fermop}
{\cal O}_\pm=-\partial^2_\sigma+\omega^2+
\frac{1-k^2\pm\sqrt{1-k^2}\sn(\sigma)\dn(\sigma)}{\cn^2(\sigma)}
\eeq

Using the periodicity properties of $\sn(\sigma)$, $\cn(\sigma)$, $\dn(\sigma)$ in \eqn{fermop} one can check that
\beq
{\cal O}_+(\sigma+2\KK)= {\cal O}_-(\sigma)\,,
\eeq
which ensures that the eigenvalue problems for the two operators are the same, so the determinants associated to the operators 
${\cal O}_+$ and ${\cal O}_-$ coincide. 

The operator $\cO_F={\cal O}_+$ (and thus ${\cal O}_-$) can be also rewritten as a single--gap Lam\'e operator
\beq
\label{fermopbis}
{\cal O}_F
=\frac{(1-k^2)(1+k_1)^2}{4}
\Big[{-}\partial_{\sigma_3}^2+\omega_3^2+2k_2^2\sn^2\big(\sigma_3+\KK_2+i\KK_2'|k_2^2\big)\Big]\,,
\quad k_2^2=\frac{4k_1}{(1+k_1)^2}\,,
\eeq
where the rescaled coordinate and frequency are now
\bal
\sigma_3&=\frac{\sqrt{1-k^2}(1+k_1)}{2}\,(\sigma+\KK), 
\qquad
&\omega_3^2&=k_2^2\left(\frac{\omega^2}{k_1(1-k^2)}-1\right).
\eal

We have rewritten all the differential operators as one--dimensional single--gap Lam\'e operators. 
With this the 1--loop effective action can be written as
\beq
\label{gammaq0}
\Gamma=-{\cal T}\int_{-\infty}^{+\infty}\frac{d\omega}{2\pi}
\ln \frac{\det^{4}{\cal O}_F}{\det^{5/2}{\cal O}_0\det {\cal O}_1\det^{1/2}{\cal O}_2}\,,
\eeq
where ${\cal T}=\int d\tau$ is the $\tau$-period. In the next subsection we 
proceed to evaluate each of these one--dimensional determinants.

\subsection{1d determinants via Gelfand-Yaglom method}
\label{app:q0-gel-yag}

We now evaluate the determinants of the Lam\'e operators written above using 
the Gelfand-Yaglom method which expresses the determinant in terms of the 
solution of an initial value problem.

This can be done since the solution of the single--gap Lam\'e eigenvalue problem
\beq\label{lamegen2}
\big[{-}\partial^2_x+2k^2\sn^2(x|k^2)\big]f(x)=\Lambda\,f(x) 
\eeq
is known in explicit form. Two independent solutions are~\cite{BradenPeriodic}
\beq\label{solslame}
y^\pm(x) = \frac{H(x\pm\alpha)}{\Theta(x)}\,e^{\mp\,x\, Z(\alpha)} \,,
\qquad
\sn(\alpha|k^2)= \frac{1}{k}\sqrt{1+k^2-\Lambda}\,.
\eeq
where the Jacobi $H$, $\Theta$ and $Z$ functions are defined in \eqn{jacobidef} in terms 
of the Jacobi $\theta$-functions.

Adapting \eqn{solslame} to the case of ${\cal O}_1$ in \eqn{O1}, one finds that two 
independent solution of the relevant differential equation are
\beq\label{solf}
y_1^\pm(\sigma)=\frac{H(\sigma_1+i\KK_1' \pm \alpha_1|k_1^2)}
{\Theta(\sigma_1+i\KK_1'|k_1^2)} \,e^{\mp Z(\alpha_1|k_1^2)(\sigma_1+ i\KK_1')}
=\frac{\vartheta_4\big(\frac{\pi\,(\sigma_1\pm\alpha_1)}{2\KK_1},q_1\big)}
{\vartheta_1\big(\frac{\pi\,\sigma_1}{2\KK_1},q_1\big)}
e^{\mp Z(\alpha_1|k_1^2)(\sigma_1+i\KK_1')\mp \frac{i\pi\alpha_1}{2\KK_1}}
\eeq
where $q_i=q(k_i^2)=\exp(-\pi\KK_i'/\KK_i)$ and
\beq\label{alpha1}
\sn(\alpha_1|k_1^2)=\frac{1}{k_1}\sqrt{1+k_1^2+\omega_1^2}
\qquad
\eeq
The solutions (\ref{solf}) diverge at the extrema of the interval $[-\KK,\KK]$ for 
the $\sigma$ variable \eqn{range}. To deal with that we use an infrared cutoff 
$\epsilon$, and the Gelfand-Yaglom theorem will be applied 
to the initial value problem with Dirichlet boundary conditions in the 
interval $-\KK+\epsilon<\sigma<\KK-\epsilon$ where $\epsilon$ is arbitrarily small. 
The linear combination 
\beq\label{u1}
u_1(\sigma)=\frac{y_1^+(-\KK+\epsilon)\,y_1^-(\sigma)-y_1^-(-\KK+\epsilon)\,y_1^+(\sigma)}{W(-\KK+\epsilon)}\,,
\eeq
where $W(\sigma)$ is the wronskian 
\beq\label{wronskian}
W(\sigma)=y_1^+(\sigma)\,\partial_\sigma y_1^-(\sigma)-\partial_\sigma y_1^+(\sigma)\,y_1^-(\sigma)
\eeq
evaluated at the regularized initial point, is a solution of the homogeneous equation with boundary conditions
\beq
u_1(-\KK+\epsilon)=0\,,
\qquad
u_1'(-\KK+\epsilon)=1\,.
\eeq
The determinant of the bosonic operator $\cO_1$ with Dirichlet boundary conditions in 
the interval $[-\KK+\epsilon,\KK-\epsilon]$ will be then given by $u_1(\KK-\epsilon)$ 
(see, for example, the discussion in Appendix C of~\cite{vali-lines}), namely
\beq\label{detO1}
\det{\cal O}_1=\frac{(k^2-1)\,\ns^2(\epsilon_1|k_1^2)-2k^2+\omega^2+1}
{\sqrt{k^2-\omega^2}\sqrt{(\omega^2-k^2+1)(-\omega^2+2k^2-1)}}\,
\sinh\big(2Z(\alpha_1)(\KK_1-\epsilon_1)+\Sigma_1\big)\,,
\eeq
where 
\beq
\Sigma_1=\ln\frac{\vartheta_4\big(\frac{\pi (\alpha_1+\epsilon)}{2\KK_1},q_1\big)}
{\vartheta_4\big(\frac{\pi (\alpha_1-\epsilon)}{2\KK_1},q_1\big)}\,,
\qquad
\epsilon_1=\sqrt{1-k^2}\,\epsilon\,.
\eeq

In a similar fashion one can work out the regularized determinants for the bosonic operator ${\cal O}_2$, obtaining
\beq
\label{detO2}
\det{\cal O}_2=
\frac{(1-k^2) (1+k_1)^2 \left(1+k_2^2+\omega _2^2-\ns^2(\epsilon_2|k_2^2\right)}{\omega\sqrt{\omega ^4+\left(2-4 k^2\right) \omega ^2+1}}\,
\sinh\big(2Z(\alpha_2)(\KK_2-\epsilon_2)+\Sigma_2\big)
\eeq
and for the fermionic fluctuations (as noticed above, it is 
$\det {\cal O}_+=\det{\cal O}_-=\det {\cal O}_F$), getting
\bal
\label{detOf}
\det{\cal O}_F&=\frac{(k_1+1)^2\left((1-k^2)(1-k_1)^2+4\omega^2\right)\dn^2(\epsilon_3|k_2^2)
-(1-k^2)(1-k_1^2)^2-4(1-k_1)^2\omega^2}
{8k_1\omega\sqrt{8(1-k^2)(k_1^2+1)\omega^2+(1-k^2)^2(1-k_1^2)^2+16\omega ^4}
\cn(\epsilon_3|k_2^2)}\\
&\quad
\times\frac{\vartheta_2\big(\frac{\pi\epsilon_3}{2\KK_2},q_2\big)}
{\vartheta_1\big(\frac{\pi\epsilon_3}{2\KK_2},q_2\big)}\,\Big(\exp\big(Z(\alpha_F)(\KK_2-2\epsilon_3)+\Sigma_F^+\big)
-\exp\big({-}Z(\alpha_F)(\KK_2-2\epsilon_3)+\Sigma_F^-\big)\Big).
\eal
The rescaled cutoffs are $\epsilon_2=(1+k_1)\sqrt{1-k^2}\,\epsilon$ and 
$\epsilon_3=\epsilon_2/2$ and the other quantities are
\beq
\begin{gathered}
\label{alpha2&f}
\sn(\alpha_2|k_2^2)=\frac{1}{k_2}\sqrt{1+k_2^2+\omega_2^2}\,,
\qquad\qquad\qquad
\sn(\alpha_F|k_2^2)=\frac{1}{k_2}\sqrt{1+k_2^2+\omega_3^2}\,,
\\
\Sigma_2=\ln\frac{\vartheta_4\big(\frac{\pi(\alpha_2+\epsilon_2)}{2\KK_2},q_2\big)}
{\vartheta_4\big(\frac{\pi(\alpha_2-\epsilon_2)}{2\KK_2},q_2\big)}\,,
\qquad
\Sigma_F^+=\ln\frac{\vartheta_4\big(\frac{\pi(\epsilon_3+\alpha_F)}{2\KK_2},q_2\big)}
{\vartheta_3\big(\frac{\pi(\epsilon_3-\alpha_F)}{2\KK_2},q_2)\big)}\,,
\qquad
\Sigma_F^-=\ln\frac{\vartheta_4\big(\frac{\pi(\epsilon_3-\alpha_F)}
{2\KK_2},q_2\big)}{\vartheta_3\big(\frac{\pi(\epsilon_3+\alpha_F)}{2\KK_2},q_2\big)}\,.
\end{gathered}
\eeq
The contribution of the massless bosons can be easily obtained via the same method 
\beq\label{detO0}
\det{\cal O}_{0}=\frac{\sinh\big(2\omega(\KK-\epsilon)\big)}{\omega}.
\eeq

\subsection{The resulting 2d determinant}
\label{app:q0-det}

With the explicit expressions \eqn{detO1}-\eqn{detO0} for the determinants of 
all our one--dimensional differential operators we would like to put them together into 
equation \eqn{gammaq0} and evaluate the one--loop effective action.

The regularization did introduce some spurious divergences that we need to 
take care of. Expanding in $\epsilon\sim0$ and retaining 
only the leading term, one gets, after some elementary manipulation,
\begin{align}
\label{O0epsilon}
\det{\cal O}_0^\epsilon
&\cong\frac{\sinh(2\KK\, \omega)}{\omega}\,,
\\
\label{O1epsilon}
\det{\cal O}_1^\epsilon
&\cong-\frac{\sinh(2\KK_1\,Z(\alpha_1))}{\epsilon^2\,\sqrt{k^2-\omega^2}\,\sqrt{(\omega^2-k^2+1)(-\omega^2+2k^2-1)}}\,,
\\
\label{O2epsilon}
\det{\cal O}_2^\epsilon
&\cong-\frac{\sinh(2\KK_2\,Z(\alpha_2))}
{\epsilon^2\,\omega\sqrt{\omega^4+(2-4k^2)\omega^2+1}}\,,
\\
\label{Ofepsilon}
\det{\cal O}_F^\epsilon
&\cong\frac{8\KK_2\sqrt{\omega_3^2+k_2^2}\sinh(\KK_2\,Z(\alpha_F))}
{\epsilon\pi(1-k^2)(k_1+1)^2\sqrt{(\omega_3^2+1)(\omega_3^2+k_2^2+1)}}
\frac{\vartheta_2(0,q_2)\,\vartheta_4\big(\frac{\pi\alpha_F}{2\KK_2},q_2\big)}
{\vartheta_1'(0,q_2)\,\vartheta_3\big(\frac{\pi\alpha_F}{2\KK_2},q_2\big)}\,.
\end{align}
where $\alpha_1$, $\alpha_2$ and $\alpha_3$ are defined in \eqn{alpha1}, \eqn{alpha2&f}.

Though we know from the analysis in Appendix~\ref{app:seeley} that the determinant should 
be finite, the integral in \eqn{gammaq0} suffers from both infrared and ultraviolet divergences. 
This is due to some of the manipulations we have done in order to get the analytic 
expressions (in particular not accounting carefully 
for boundary counter--terms). We expect that subtracting 
the divergent terms will lead to the correct finite expressions.

The infrared divergences are from small $\epsilon$ where
\beq
4\ln\det{\cal O}_F-\frac{5}{2}\ln\det {\cal O}_0-\ln\det{\cal O}_1-\frac{1}{2}\ln\det{\cal O}_2
\sim\ln\frac{1}{\epsilon}\,,
\eeq
The UV divergences come from large $\omega$, 
where the general structure of the expansion can nicely be obtained 
in terms of the analytically known eigenvalues of the single gap Lam\'e potentials (\ref{O1}), 
(\ref{O2}) and (\ref{fermopbis}).%
\footnote{See the analysis in~\cite{bdfpt}.} 
It can also be found directly by use of the representations (\ref{intzeta}) and (\ref{zed}), 
where after some manipulation one finds
\bal\label{largeomega}
\ln\det{\cal O}_0&=2\KK\omega-\ln\omega-\ln2\,,
\\
\ln\det{\cal O}_1
&=2\KK\omega-3\ln\omega-\ln2 +\frac{2}{\omega}\,\frac{\KK_1-\EE_1}{\sqrt{1-k_1^2}}
+O(\omega^{-3})\,,
\\ 
\ln\det{\cal O}_2
&=2\KK\omega-3\ln\omega-\ln2 +\frac{4}{\omega}\,\frac{\KK_1-\EE_1}{\sqrt{1-k_1^2}}
+O(\omega^{-3})\,,
\\
\ln\det{\cal O}_F
&=2\KK\omega-2\ln\omega-\ln2 +\frac{1}{\omega}\,\frac{\KK_1-\EE_1}{\sqrt{1-k_1^2}}
+O(\omega^{-3})\,,
\eal
where $\EE_1=\EE(k_1^2)$ and in the expansion of the fermionic determinant we have used 
\beq
\ln\left[\frac{\vartheta_2(0,q_2)\,\vartheta_4\big(\frac{\pi \alpha_F}{2 \KK_2},q_2\big)}
{\vartheta_1^{\prime }(0,q_2)\,\vartheta_3\big(\frac{\pi \alpha_F}{2 \KK_2},q_2\big)}\right]
=-\ln\omega-\ln\frac{4\KK}{\pi }+\frac{i \pi }{2}+O(\omega^{-3})\,.
\eeq
Therefore for large $\omega$
\beq
\label{uvdiv}
4\ln\det{\cal O}_F-\frac{5}{2}\ln\det {\cal O}_0-\ln\det{\cal O}_1-\frac{1}{2}\ln\det{\cal O}_2
\sim\ln\frac{1}{\omega}\,,
\eeq

Explicitly subtracting the divergences we find the finite expression for the regularized 
1--loop effective action%
\footnote{Such regularization is often attributed to the need to subtract the contribution 
of a straight line. This is not so, since when defined properly, the Wilson loop is a finite 
observable (apart for the infinite extension of the lines). In fact, if we consider the 
divergences that arise when applying our prescription to the straight line (by taking 
the $k\to0$ limit of our expressions), we will find different values of both ${\cal T}$ and 
$\epsilon$, and the divergences will {\em not} cancel. Instead, the divergences 
which are a regularization artifact should be removed by standard renormalization. }
\beq
\label{gammareg}
\Gamma_\text{reg}=-\frac{{\cal T}}{2}\lim_{\epsilon\to0}\int_{-\infty}^{+\infty}\frac{d\omega}{2\pi}
\ln\frac{\epsilon^2\omega^2\det^{8}{\cal O}_F^\epsilon}
{\det^{5}{\cal O}_0^\epsilon\det^2{\cal O}_1^\epsilon\det{\cal O}_2^\epsilon}\,,
\eeq
with the explicit expressions for the 1d determinants given in \eqn{O0epsilon}-\eqn{Ofepsilon}. 
This is the exact one--loop contribution to the string partition function from which we can 
read off the one--loop correction to the effective potential
\beq
V^{(1)}_{AdS}(\phi,0)=\frac{1}{T}\Gamma_\text{reg}\,.
\eeq
The integral in \eqn{gammareg} can then be evaluated numerically, or as we do in the 
next subsection, expanded in a power series for small $k$ and evaluated analytically.

\subsection{Expansion for small $\phi$}
\label{app:q0-expand}

The small $\phi$ expansion is realized sending $k\to 0$ (equivalently $p\to\infty$) in the 
expressions or the determinants (\ref{O0epsilon})-(\ref{Ofepsilon}).
An efficient way to proceed, considering for example the determinant for the operator 
${\cal O}_1$, is to transform as follows
\beq\label{shift}
\alpha_1=\beta_1+\KK_1+i\KK_1'
\eeq
which allows to identify the imaginary part of the argument of the hyperbolic function
\beq\label{eq:newbeta}
2\KK_1Z(\alpha_1)=2\KK_1 Z(\beta_1)-2\KK_1
\frac{\sn(\beta_1|k_1^2)\dn(\beta_1|k_1^2)}{\cn(\beta_1|k_1^2)}-i\pi
\eeq
In applying this approach to the fermionic determinant, one notices that a shift 
analog to (\ref{shift}) changes the $\sinh$ in $\cosh$. One can then first compute 
the $k\to0$ expansion of $\partial Z(\alpha_i|k^2)/\partial\omega$ (using the integral 
representation (\ref{intzeta})) where the dependence of $Z$ on $\omega$ is via 
$\alpha$, and then perform an indefinite integration over $\omega$. 

From examining the expansion of the determinants at small $k$ we find the form
\beq
\det \cO_i =\sum_{l=0}^\infty D_i^{(l)}k^{2l}\,, 
\qquad i=0,1,2,F\,,
\eeq
where each $D_i^{(l)}$ is a rational function in $\omega$ times $\sinh(\pi\omega)$ and 
$\cosh(\pi\omega)$. The first few are
\begin{align}
&D_0^{(0)}=\frac{\sinh(\pi\omega)}{\omega}\,,
\qquad
D_0^{(2)}=\frac{\pi}{4}\cosh(\pi\omega)\,,
\qquad
D_0^{(4)}=\frac{\pi^2\omega}{32}\sinh(\pi\omega)+\frac{9\pi}{64}\cosh(\pi\omega)\,,
\nonumber\\
&D_1^{(0)}=\frac{\sinh(\pi\omega)}{\omega(\omega^2+1)}\,,
\qquad
D_1^{(2)}=
\frac{\pi(\omega^2-2)\cosh(\pi\omega)}{4\omega^2(\omega^2+1)}
+\frac{(4\omega^2+1)\sinh(\pi\omega)}{2\omega^3(\omega^2+1)^2}\,,
\nonumber\\*
&D_1^{(4)}=
\frac{\pi(9\omega^6+37\omega^4-72\omega^2-24)\cosh(\pi\omega)}{64\omega^4(\omega^2+1)^2}
+\left(\frac{28\omega^4+12\omega^2+3}{8\omega^5(\omega^2+1)^3}
+\frac{\pi^2(\omega^2-2)^2}{32\omega^3(\omega^2+1)}\right)\sinh(\pi\omega)\,,
\nonumber\\
&D_2^{(0)}=\frac{\sinh(\pi\omega)}{\omega(\omega^2+1)}\,,
\qquad
D_2^{(2)}=\frac{2\omega\sinh(\pi\omega)}{(\omega^2+1)^3}
+\frac{\pi(\omega^2-3)\cosh(\pi\omega)}{4(\omega^2+1)^2}\,,
\nonumber\\*
&D_2^{(4)}=\left(\frac{\pi^2(\omega^2-3)^2\omega}{32(\omega^2+1)^3}
+\frac{6\omega^3}{(\omega^2+1)^5}\right)\sinh(\pi\omega)
+\frac{3\pi(3\omega^6+17\omega^4-55\omega^2-5)\cosh(\pi
\omega)}{64(\omega^2+1)^4}\,,
\nonumber\\
&D_F^{(0)}=\frac{4\cosh(\pi\omega)}{4\omega^2+1}\,,
\qquad
D_F^{(2)}=\frac{\pi(4\omega^2-3)\omega\sinh(\pi\omega)}{(4\omega^2+1)^2}+\frac{32\omega^2\cosh
(\pi\omega)}{(4\omega^2+1)^3}\,,
\nonumber\\*
&D_F^{(4)}=\frac{3\pi\omega(192\omega^6+272\omega^4-220\omega^2-5)\sinh(\pi\omega)}{16(4\omega^2+1)^4}
\nonumber\\*
&\qquad\quad
+\left(\frac{\pi^2(3-4\omega^2)^2\omega^2}{8(4\omega^2+1)^3}
+\frac{384\omega^4}{(4\omega^2+1)^5}\right)\cosh(\pi\omega)\,.
\end{align}

The zeroth order contribution to the regularized effective action (\ref{gammareg}) in this limit reads then
\bal\label{zero}
\frac{\Gamma_\text{reg}^{(0)}}{\cal T}
&=-\frac{1}{4\pi}\int_{-\infty}^{+\infty}d\omega
\ln\left[\frac{\big(D_F^{(0)}\big)^8}{\big(D_1^{(0)}\big)^2\big(D_2^{(0)}\big)\big(D_0^{(0)}\big)^5}\right]
\\&
=-\frac{1}{4\pi}\int_{-\infty}^{+\infty}d\omega
\ln\left[\frac{2^{16}\,\omega^{10}(\omega^2+1)^3\coth^8(\pi\omega)}{(4\omega^2+1)^8}\right]=0
\eal
At order $k^2$ the result is
\beq\label{two}
\frac{\Gamma_\text{reg}^{(2)}}{\cal T}=-\frac{1}{4\pi}\int_{-\infty}^{+\infty}d\omega
\left[8\frac{D_F^{(2)}}{D_F^{(0)}}
-2\frac{D_1^{(2)}}{D_1^{(0)}}-\frac{D_2^{(2)}}{D_2^{(0)}}-5\frac{D_0^{(2)}}{D_0^{(0)}}\right]
=\frac{3}{8}\,.
\eeq
At order $k^4$ one finds
\bal\label{four}
\frac{\Gamma_\text{reg}^{(4)}}{\cal T}=-\frac{1}{4\pi}&\int_{-\infty}^{+\infty}
d\omega\Bigg[
8\left(\frac{D_F^{(4)}}{D_F^{(0)}}-\frac{\big(D_F^{(2)}\big)^2}{2\big(D_F^{(0)}\big)^2}\right)
-2\left(\frac{D_1^{(4)}}{D_1^{(0)}}-\frac{\big(D_1^{(2)}\big)^2}{2\big(D_1^{(0)}\big)^2}\right)
\\&\quad
-\left(\frac{D_2^{(4)}}{D_2^{(0)}}-\frac{\big(D_2^{(2)}\big)^2}{2\big(D_2^{(0)}\big)^2}\right)
-5\left(\frac{D_0^{(4)}}{D_0^{(0)}}-\frac{\big(D_0^{(2)}\big)^2}{2\big(D_0^{(0)}\big)^2}\right)
\Bigg]
=\frac{29}{128}-\frac{3\,\zeta(3)}{16}\,.
\eal 
Proceeding in a similar way, one finds at orders $k^6$ and $k^8$
\bal\label{sixeight}
\frac{\Gamma_\text{reg}^{(6)}}{\cal T}  & =\frac{121}{512}-\frac{15 \,\zeta(3)}{128}-\frac{15 \,\zeta(5)}{128}\,,\\
\frac{\Gamma_\text{reg}^{(8)}}{\cal T} & 
=\frac{9669}{32768}-\frac{229\, \zeta(3)}{2048}-\frac{115\, \zeta(5)}{1024}-\frac{315 \,\zeta(7)}{4096}\,.
\eal
To perform these integrals one examines the behavior of the integrand at large imaginary argument, 
which asymptotes to a polynomial in $\omega$ and hyperbolic functions. After subtracting this 
asymptotic expression, whose integral can be found in standard tables, 
the remainder can be integrated by closing the contour around the upper 
half plane taking care of the poles from the rational functions at $i/2$ and $i$ and the poles 
from the hyperbolic functions at all integer or half-integer imaginary values.

In terms of a $p\to\infty$ expansion, the 1--loop energy is then written as
\bal
\label{V1pexp}
V^{(1)}_{AdS}=&\,\frac{1}{T}\,\Gamma_\text{reg}
=\frac{1}{T}\Big[\Gamma_\text{reg}^{(0)}+k^2\Gamma_\text{reg}^{(2)}
+k^4\Gamma_\text{reg}^{(4)}+k^6\Gamma_\text{reg}^{(6)}+k^8\Gamma_\text{reg}^{(8)}
+O(k^{10})\Big]\\
=&\,\frac{3}{8}\frac{1}{p^2}-\left(\frac{67}{128}+\frac{3\,\zeta(3)}{16}\right)\frac{1}{p^4}
+\left(\frac{597}{512}+\frac{105\,\zeta(3)}{128}-\frac{15\,\zeta(5)}{128}\right)\frac{1}{p^6}\\
&\,
{-}\left(\frac{101563}{32768}+\frac{6565\,\zeta(3)}{2048}-\frac{845\,\zeta(5)}{1024}
+\frac{315\,\zeta(7)}{4096}
\right)
\frac{1}{p^8}
+O(p^{-10})\,,
\eal
where we have used (\ref{rescaling}) and (\ref{b}) ($q=0$) which give, in this limit,
\beq
{\cal T}/T=\frac{1}{\sqrt{1-2 k^2}}
=\frac{(p^2+4)^{1/4}}{\sqrt{p}}\sim 1+p^{-2}-\textstyle{\frac{3}{2}} \,p^{-4}+\frac{7}{2}\,p^{-6}-\frac{77}{8}\,
   p^{-8}+O\,(p^{-10})\,.
\eeq

\section{One--loop determinant for $\phi=0$}
\label{app:phi=0}

In this appendix we study another special case of the general fluctuation operator 
derived in Appendix~\ref{app:fluc}. We consider the fluctuation about the minimal 
surface which is entirely within an $AdS_2\times\bS^1$ subspace of 
$AdS_5\times\bS^5$. This configuration is achieved for $p\to\infty$ while keeping 
$k$ finite, such that $q/p=ik/\sqrt{1-k^2}$ and $b/p=1/\sqrt{1-k^2}$. 
Now $k$ is imaginary and can take arbitrary values along the imaginary 
axis.

The expressions we write below are valid (and can be evaluated reliably 
in Mathematica) for $-1<k^2<0$ which corresponds to 
$|\theta|<\Gamma(\frac{1}{4})^2/(2\sqrt{2\pi})\sim2.62206$. 
Some care is required to analytically continue beyond that value. 
Note that in the string solution $\theta$ is not restricted to be bound by $\pm\pi$. 
In our $AdS_3\times S^1$ ansatz, $\vartheta$ \eqn{metric1} parameterizes a 
noncontractible cycle, so $\theta$ can take any real value. Solutions with 
$|\theta|>\pi$ are unstable in the full space and are subdominant saddle points.

In this limit the fluctuation field $\zeta_8$ simplifies to 
$\sinh\rho\,\delta\varphi$ and has the same action as 
$\zeta_1$ and $\zeta_2$.
In the quadratic Lagrangean \eqn{bosfluct} the mass parameters become
\beq
M_{11}=M_{22}=M_{88}=2+\frac{k^2}{\sqrt{g}}\,,\qquad
M_{33}=\cdots=M_{66}=\frac{k^2}{\sqrt{g}}\,,\qquad
M_{77}=R^{(2)}+2+\frac{k^2}{\sqrt{g}}\,,
\eeq
where $R^{(2)}$ is the scalar curvature of the induced metric \eqn{curvature} and 
$A=0$.

In this limit the parameter $\beta$ in the mass term (\ref{fermionmass}) of the fermionic 
fluctuation operator \eqn{2d-fermi} goes to $\pi/2$ and we find
\beq
M_F= i\sqrt{1+\frac{k^2 }{\sqrt{g}}}\,\Gamma_{127}
= \sqrt{1+\frac{k^2 }{\sqrt{g}}}\,\gamma_3
= \frac{\dn(\sigma| k^2)}{\sqrt{1 - k^2}}\,\gamma_3\,.
\eeq 

Thus the analog of the partition function (\ref{Z-q=0}) is in this limit
\beq
\label{Z-p=inf}
Z=\frac{\det^4\big({-}i\gamma^i\,\hat\nabla_i+\sqrt{1+\frac{k^2 }{\sqrt{g}}}\,\gamma_3\big)}
{\det^{2}\big({-}\nabla^2+\frac{k^2}{\sqrt g}\big)
\det^{3/2}\big({-}\nabla^2+2+\frac{k^2}{\sqrt g}\big)
\det^{1/2}\big({-}\nabla^2+R^{(2)}+2+\frac{k^2}{\sqrt g}\big)}\,.
\eeq

After rescaling all the fluctuation operators by $\sqrt{g}$ and Fourier transforming 
$\partial_\tau\to i\omega$ one finds
\begin{align}
\label{opbos0-2}
\widetilde\cO_0=\sqrt{g}\left(-\nabla^2+\frac{k^2}{\sqrt{g}}\right)
&=-\partial_\sigma^2+\omega^2+k^2\,,\\
\label{opbos1-2}
\widetilde\cO_1=\sqrt{g}\left(-\nabla^2+2+\frac{k^2}{\sqrt{g}}\right)
&=-\partial_\sigma^2+\omega^2+\frac{2(1-k^2)}{\cn^2(\sigma)}+k^2\,,\\
\label{opbos2-2}
\widetilde\cO_2=\sqrt{g}\left(-\nabla^2+R^{(2)}+2+\frac{k^2}{\sqrt{g}}\right)
&=-\partial_\sigma^2+\omega^2+2k^2\sn^2(\sigma)-k^2\,.
\end{align}
Note that for two of the bosonic operators, $\widetilde\cO_0$ and $\widetilde\cO_1$ 
we have the same formal expressions as in the $\theta=0$ case in 
Appendix~\ref{app:theta=0},  \eqn{opbos0} and \eqn{opbos1}, with a shift 
by $k^2$. The operator $\widetilde\cO_2$ did not appear before, 
but it too is of the Lam\'e type.

Simplifying the fermionic operator is very similar to the $\theta=0$ case. Here the operator reads explicitly
\beq\label{fermopphi0}
\widetilde D_F=-i\gamma^i\,\hat\nabla_i+\frac{\dn(\sigma)}{\sqrt{1 - k^2}}\gamma_3
=\frac{\cn(\sigma)}{\sqrt{1-k^2}}
\left[-i\left(\partial_\sigma+\frac{\sn(\sigma)\dn(\sigma)}{2\cn(\sigma)}\right)\tau_1
-\omega\,\tau_2+\frac{\dn({\sigma})}{\cn(\sigma)}\,\tau_3\right],
\eeq
Squaring and diagonalizing as in (\ref{diag}), one gets
\beq
\widetilde{\cal O}_\pm=-\partial_\sigma^2+\omega^2+\frac{1\pm k^2 \sn(\sigma)}{1\pm \sn(\sigma)}\,.
\eeq
Again, the periodicity $\widetilde{\cal O}_+(\sigma+\KK)=\widetilde{\cal O}_-(\sigma)$ allows 
us to deal with only one operator (say ${\widetilde{\cal O}}_F=\widetilde{\cal O}_+$), which can 
be written as a Lam\'e  operator
\beq
\widetilde{ {\cal O}}_F=\Big(\frac{1+k}{2}\Big)^2 \Big[{-}\partial_{\sigma_4}^2+\omega_4^2+2\bar k^2\sn^2\Big(\sigma_4-\textstyle{\frac{3}{2}}\KK_4+i\KK_4'\big|k^2_4\Big)-k_4^2\Big],
\eeq
where $\sigma_4=(1+k)\sigma/2$, $\omega_4^2=4\,\omega^2/(1+k)^2$ and
\beq
k^2_4=\frac{4k}{(1+k)^2}\,,
\qquad
\KK_4=\KK(k^2_4)\,,
\qquad
\KK_4'=\KK(1-k^2_4)\,.
\eeq

The rest of the calculation goes through as before.  For the operators $\widetilde\cO_0$ 
and $\widetilde\cO_1$ the expressions for the regularized determinants can be just read 
off from (\ref{O0epsilon}) and (\ref{O1epsilon}) via the shift $\omega^2\to\omega^2+k^2$. 
In the case of $\widetilde{ {\cal O}}_2$ and $\widetilde{ {\cal O}}_F$ one proceeds with 
the Gelfand-Yaglom method as in Appendix~\ref{app:q0-gel-yag}, obtaining
\bal
\label{detO2bis}
\det{\widetilde{\cal O}}_2=&\,
{-}\frac{\omega ^2 \text{dn}(\epsilon|k^2)^2-k^2+1}
{\omega\sqrt{(\omega ^2+1)(-k^2+\omega ^2+1)}\,\dn(\epsilon|k^2)^2}
\sinh\big(2Z(\widetilde\alpha_2)(\KK-\epsilon)+\widetilde{\Sigma}_2\big)
\eal
where $\widetilde{\alpha}_2$ and $\widetilde{\Sigma}_2$ are defined via
\beq
\sn(\widetilde{\alpha}_2|k^2)=\sqrt{\frac{1+\omega^2}{k^2}}\,,
\qquad
\Sigma_2=\ln\frac{\vartheta_2\big(\frac{\pi(\epsilon+\bar\alpha_2)}{2\KK},q\big)}
{\vartheta_2\big(\frac{\pi(\epsilon-\bar\alpha_2)}{2\KK},q\big)}\,,
\eeq
and 
\bal
\label{detOfbis}
\det{\widetilde{\cal O}_F}=&\,
\frac{(1+k)^2+4\omega^2-(1+k)^2\ns^2(\epsilon_4|k^2_4)}
{2\,\omega\sqrt{(1-k)^2+4\omega^2}\sqrt{(k+1)^2+4\omega^2}}\,
\frac{\theta_1\big(\frac{\pi\bar{\epsilon}}{2\KK_4},q_4\big)}
{\theta_2\big(\frac{\pi\bar\epsilon}{2\KK_4},q_4\big)}\\
&\,
\times\Big(\exp\big(Z(\widetilde\alpha_F)(\KK_4-2\epsilon_4)+\widetilde{\Sigma}_F^+\big)
-\exp\big({-}Z(\widetilde\alpha_F)(\KK_4-2\epsilon_4)+\widetilde{\Sigma}_F^-\big)\Big).
\eal
where $\epsilon_4=(1+k)\epsilon/2$, $q_4=\exp(-\pi\KK_4'/\KK_4)$ and
\beq
\sn(\widetilde\alpha_F|k^2_4)=\frac{1}{2} \sqrt{\frac{(k+1)^2+4 \omega ^2}{k}},
\qquad
\widetilde{\Sigma}_F^+=\ln\frac{\vartheta_3
\big(\frac{\pi(\epsilon_4+\widetilde\alpha_f)}{2\KK_4},q_4\big)}
{\vartheta_4\big(\frac{\pi(\epsilon_4-\widetilde\alpha_f)}{2\KK_4},q_4\big)},
\qquad
\widetilde{\Sigma}_F^-=\ln\frac{\vartheta_3
\big(\frac{\pi(\epsilon_4-\widetilde\alpha_f)}{2\KK_4},q_4\big)}
{\vartheta_4\big(\frac{\pi(\epsilon_4+\widetilde\alpha_f)} {2\KK_4},q_4\big)}.
\eeq

Expanding in $\epsilon\sim0$ one obtains 
\begin{align}
\label{O0epsilonbis}
\det{\widetilde{\cal O}}_0^\epsilon
&\cong\frac{\sinh \left(2 \KK\, \sqrt{k^2+\omega ^2}\right)}{\sqrt{k^2+\omega ^2}}\,,
\\
\label{O1epsilonbis}
\det{\widetilde{\cal O}}_1^\epsilon
&\cong\frac{\sinh (2\KK_1 Z(\widetilde\alpha_1))}{\epsilon ^2 \sqrt{\omega ^2 (\omega ^2+1)(1-k^2+\omega ^2)}}\,,
\\
\label{O2epsilonbis}
\det{\widetilde{\cal O}}_2^\epsilon
&\cong-\frac{\sqrt{1-k^2+\omega ^2}}{\omega\,\sqrt{1+\omega^2}}\sinh (2 \KK Z(\widetilde\alpha_2))\,,
\\
\label{Ofepsilonbis}
\det{\widetilde{\cal O}}_F^\epsilon
&\cong-\frac{\pi(k+1)\sinh(\KK_4\,Z(\widetilde\alpha_F))}
{\epsilon\,\KK_4\,\omega\sqrt{k^4+2k^2(4\omega^2-1)+(4\omega^2+1)^2}}
\,\frac{\vartheta_1'(0,q_4)
\vartheta_3\big(\frac{\pi\widetilde\alpha_F}{2\KK_4},q_4\big)}
{\vartheta_2(0,q_4)
\vartheta_4\big(\frac{\pi\widetilde\alpha_F}{2\KK_4},q_4\big)}\,.
\end{align}
As in the previous case there are infrared divergences from small $\epsilon$. 
To see the ultraviolet behavior we expand these expressions for large $\omega$ to find
\bal
\label{largeomegabis}
\ln\det{\widetilde{\cal O}}_0
&=2\KK\omega-\ln\omega-\ln2+\frac{k^2\,\KK}{\omega}+O(\omega^{-3})\,,
\\
\ln\det{\widetilde{\cal O}}_1
&=2\KK\omega-3\ln\omega-\ln2+\frac{(2-k^2)\KK-2\sqrt{1-k^2}\,\EE_1}{\omega}+O(\omega^{-3})\,,
\\
\ln\det{\widetilde{\cal O}}_2
&=2\KK\omega-\ln\omega-\ln2+\frac{(2-k^2)\KK-2\EE}{\omega}+O(\omega^{-3})\,,
\\
\ln\det{\widetilde{\cal O}}_F
&=2\KK\omega-2\ln\omega-\ln2+\frac{(k^2+1)\KK-(k+1)\EE_4}{2\omega}+O(\omega^{-3})\,,
\eal
where $\EE_4=\EE(k^2_4)$.
Using elliptic integral identities, the weighted sum of these expressions gives 
$-\ln\omega$, exactly like in \eqn{uvdiv}.

With the extra $\epsilon^2\omega^2$ to cancel the IR and UV divergences, 
the analog of (\ref{gammareg}) reads here
\beq\label{gammaregbis}
\widetilde{\Gamma}_\text{reg}
=-\frac{{\cal T}}{2}\lim_{\epsilon\to0}\int_{-\infty}^{+\infty}\frac{d\omega}{2\pi}
\ln\frac{\epsilon^2\omega^2\det^{8}{\widetilde{\cal O}}_F^\epsilon}
{\det^{4}{\widetilde{\cal O}}_0^\epsilon\det^3{\widetilde{\cal O}}_1^\epsilon\det{\widetilde{\cal O}}_2^\epsilon}\,.
\eeq
And the one--loop correction to the effective potential is given by
\beq
V^{(1)}_{AdS}(0,\theta)=\frac{1}{T}\Gamma_\text{reg}\,,
\eeq
which can be evaluated numerically, or when expanded in small $k$, also analytically, 
as we do now.

\subsection{Expansion for small $\theta$}
\label{app:pinf-expand}

The small $\theta$ expansion can be carried out in total analogy with the expansion of 
Section~\ref{app:q0-expand}, since expanding around the BPS configuration coincides with an 
expansion in small $k$.

One finds 
\beq
\det \widetilde \cO_i =\sum_{l=0}^\infty \widetilde D_i^{(l)}k^{2l}\,, 
\qquad i=0,1,2,F\,,
\eeq
where the first terms in the series read
\begin{align}
&\widetilde D_0^{(0)}=\frac{\sinh(\pi\omega)}{\omega}\,,
\qquad
\widetilde D_0^{(2)}=\frac{\pi(\omega^2+2)\cosh(\pi\omega)}{4\omega^2}-\frac{\sinh(\pi\omega)}{2\omega^3}\,,
\nonumber\\*
&\widetilde D_0^{(4)}=\frac{3\pi(3\omega^4-8)\cosh(\pi\omega)}{64\omega^4}+\frac{(\pi^2\omega^2(\omega^2+2)^2+12)\sinh(\pi\omega)}{32\omega^5}\,,
\nonumber\\
&\widetilde D_1^{(0)}=\frac{\sinh(\pi\omega)}{\omega(\omega^2+1)}\,,
\qquad
\widetilde D_1^{(2)}=
\frac{\pi\cosh(\pi\omega)}{4(\omega^2+1)}
+\frac{\sinh(\pi\omega)}{2\omega(\omega^2+1)^2}\,,
\nonumber\\*
&\widetilde D_1^{(4)}=
\frac{3\pi(3\omega^3+7)\cosh(\pi\omega)}{64(\omega^2+1)^2}
+\left(\frac{\pi^2\omega}{32(\omega^2+1)}+\frac{3}{8\omega(\omega^2+1)^3}\right)\sinh(\pi\omega)\,,
\nonumber\\
&\widetilde D_2^{(0)}=\frac{\sinh(\pi\omega)}{\omega}\,,
\qquad
\widetilde D_2^{(2)}=\frac{\pi}{4}\cosh(\pi\omega)-\frac{\sinh(\pi\omega)}{2\omega(\omega^2+1)}\,,
\\*
&\widetilde D_2^{(4)}=\frac{\pi(9\omega^2+5)\cosh(\pi\omega)}{64(\omega^2+1)}
+\left(\frac{\pi^2\omega}{32}-\frac{1}{8\omega(\omega^2+1)^2}\right)\sinh(\pi\omega)\,,
\nonumber\\
&\widetilde D_F^{(0)}=\frac{4\cosh(\pi\omega)}{4\omega^2+1}\,,
\qquad
\widetilde D_F^{(2)}=\frac{4(1-4\omega^2)\cosh(\pi\omega)}{(4\omega^2+1)^3}
+\frac{\pi\omega(4\omega^2+5)\sinh(\pi\omega)}{(4\omega^2+1)^2}\,,
\nonumber\\*
&\widetilde D_F^{(4)}=\frac{\pi\omega(576\omega^6+560\omega^4-20\omega^2+161)\sinh(\pi\omega)}{16\,(4\omega^2+1)^4}\nonumber\\*
&\qquad\quad
+\left(\frac{\pi^2\omega^2(4\omega^2+5)^2}{8(4\omega^2+1)^3}
+\frac{4(16\omega^4-16\omega^2+1)}{(4\omega^2+1)^5}
\right)\cosh(\pi\omega)\,.
\nonumber
\end{align}
The resulting  first contributions to the regularized effective action (\ref{gammaregbis}), formally defined as in (\ref{zero})-(\ref{four}), are evaluated by the same means and read 
\begin{align}
\label{resultsbis}
&\frac{\widetilde\Gamma_\text{reg}^{(0)}}{\cal T}=0\,,
\qquad\qquad
\frac{\widetilde\Gamma_\text{reg}^{(2)}}{\cal T}=\frac{3}{8}\,,
\qquad&
\frac{\widetilde\Gamma_\text{reg}^{(4)}}{\cal T}&=\frac{5}{128}-\frac{3\zeta(3)}{16}\,,
\\
&\frac{\widetilde\Gamma_\text{reg}^{(6)}}{\cal T}
=\frac{3}{512}-\frac{15\,\zeta(3)}{128}+\frac{15\,\zeta(5)}{128}\,,
\qquad
&
\frac{\widetilde\Gamma_\text{reg}^{(8)}}{\cal T}
&=-\frac{59}{32768}-\frac{173\,\zeta(3)}{2048}+\frac{145\,\zeta(5)}{1024}-\frac{315\,\zeta(7)}{4096}\,.
\nonumber
\end{align}
The 1--loop energy is then written as 
\bal
\label{V1pexpbis}
V^{(1)}_{AdS}=&\,\frac{1}{T}\,\widetilde\Gamma_\text{reg}
=\frac{1}{T}\Big[\widetilde\Gamma_\text{reg}^{(0)}+k^2\widetilde\Gamma_\text{reg}^{(2)}+k^4\widetilde\Gamma_\text{reg}^{(4)}+k^6\widetilde\Gamma_\text{reg}^{(6)}+k^8\widetilde\Gamma_\text{reg}^{(8)}+O(k^{10})\Big]
\\
=&\,{-}\frac{3}{8}\frac{q^2}{p^2}-\left(\frac{19}{128}+\frac{3\,\zeta(3)}{16}\right)\frac{q^4}{p^4}
-\left(\frac{45}{512}+\frac{21\,\zeta(3)}{128}+\frac{15\,\zeta(5)}{128}\right)\frac{q^6}{p^6}\\
&\,
{-}\left(\frac{1979}{32768}+\frac{293\,\zeta(3)}{2048}
+\frac{155\,\zeta(5)}{1024}+\frac{315\,\zeta(7)}{4096}\right)\frac{q^8}{p^8}
+{\cal O}((q/p)^{10})\,,
\eal
where, using (\ref{rescaling}) and (\ref{b}) in this limit, it is
\beq
{\cal T}/T=\frac{1}{\sqrt{1-k^2}}\sim 1-\frac{q^2}{2 p^2}-\frac{q^4}{8\,p^4}
-\frac{q^4}{16\,p^4}-\frac{5\,q^8}{128\,p^8}+{\cal O}((q/p)^{10})\,.
\eeq

\section{Elliptic functions}
\label{app:funcs}

The \emph{incomplete elliptic integrals} of the first, second and third kind are defined via 
\beq
\begin{gathered}
\label{incomplete}
F(x|k^2)=\int_0^{x}d\theta(1-k^2\sin^2\theta)^{-1/2},
\qquad
E(x|k^2) =\int_0^{x}d\theta (1-k^2\sin^2\theta)^{1/2}
\\
\Pi(\ell^2;x|k^2)=\int_0^{x}\frac{d\theta}{(1-\ell^2\sin^2\theta)\,(1-k^2\sin^2\theta)^{-1/2}},
\end{gathered}
\eeq
where $k^2$ is their modulus and $\ell^2$ is the characteristic. 

The corresponding \emph{complete} elliptic integrals are given by
\beq
\KK=\KK(k^2)=F({\textstyle\frac{\pi}{2}}|k^2)\,,
\qquad
\EE=\EE(k^2)=E({\textstyle\frac{\pi}{2}}|k^2)\,,\qquad
\Pi(\ell^2|k^2)=\Pi(\ell^2;{\textstyle\frac{\pi}{2}}|k^2)\,.
\eeq
Defining the \emph{Jacobi amplitude} as
\beq
\vartheta=\am (u|k^2),\qquad\text{where}\qquad
u=F(\arcsin\vartheta|k^2)
\eeq
the \emph{Jacobi elliptic functions} $\sn,\cn,\dn$ are defined by
\beq
\sn(u|k^2)=\sin \vartheta\,,\qquad
\cn(u|k^2)=\cos \vartheta\,,\qquad
\dn(u|k^2)=\sqrt{1-k^2\sin^2 \vartheta}
\eeq
and, for example, $\ns(u|k^2)=1/\sn(u|k^2)$, 
$\sd(u|k^2)={\sn(u|k^2)}/{\dn(u|k^2)}$, and $\cd(u|k^2)={\cn(u|k^2)}/{\dn(u|k^2)}$.

Useful relations between the squares of the functions are
\beq
\begin{gathered}
-\dn^2(u|k^2)+k'^2=-k^2\cn^2(u|k^2)=k^2\sn^2(u|k^2)-k^2\\
-k'^2\nd(u|k^2)+k'^2=-k^2k'^2\sd^2(u|k^2)=k^2\cd(u|k^2)-k^2.
\end{gathered}
\eeq
where $k'^2=1-k^2$.

The Jacobi $H$, $\Theta$ and $Z$ functions are defined as follows in terms of the Jacobi $\theta$ functions
\beq\label{jacobidef}
H(u|k^2) = \vartheta_1\left(\frac{\pi\,u}{2\,\KK}, q\right), \qquad
\Theta(u|k^2) = \vartheta_4\left(\frac{\pi\,u}{2\,\KK}, q\right),\qquad
Z(u|k^2) = \frac{\pi}{2\,\KK}\,\frac{\vartheta_4'(\frac{\pi\,u}{2\,\KK}, q)}
{\vartheta_4(\frac{\pi\,u}{2\,\KK}, q)}
\eeq
where $q=q (k^2) = \exp(-\pi\frac{\KK'}{\KK})$.

Useful representations for $Z(u|k^2)$ are the integral representation
\beq\label{intzeta}
Z(\mathrm{sn}^{-1}(y|k^2)|k^2)=\int_0^y\mathrm{d}t\left[\sqrt{\frac{1-k^2t^2}{1-t^2}}-\frac{\EE(k^2)}{\KK(k^2)}
\frac{1}{\sqrt{(1-t^2)(1-k^2t^2)}}\right]
\eeq
and
\beq
{\mathbb Z}(\alpha|k^2)=\int_0^\alpha du \dn^2(u|k^2) - \frac{{\mathbb E}(k^2)}{{\mathbb K}(k^2)}
\, \alpha\ .
\label{zed}
\eeq
valid for $0<\alpha<\KK$.

\bibliography{refs}
\end{document}